\documentclass[11pt]{article}
\usepackage{geometry}
\usepackage{amsmath}
\usepackage{amssymb}
\usepackage{latexsym}
\usepackage{epsfig}

\usepackage[vcentermath]{youngtab}

\usepackage{amsmath,amssymb,amsbsy}
\usepackage{latexsym,layout,amsfonts,amssymb,fontenc,xcolor,amsthm,multirow,amsfonts,bm,textcomp}
\usepackage{hyperref}
\usepackage{empheq}
\usepackage{accents}

\newsavebox{\uuunit}
\sbox{\uuunit}
    {\setlength{\unitlength}{0.825em}
     \begin{picture}(0.6,0.7)
        \thinlines
        \put(0,0){\line(1,0){0.5}}
        \put(0.15,0){\line(0,1){0.7}}
        \put(0.35,0){\line(0,1){0.8}}
       \multiput(0.3,0.8)(-0.04,-0.02){12}{\rule{0.5pt}{0.5pt}}
     \end {picture}}

\def\2{\frac12}
\def\4{\frac14}

\def\equationautorefname~#1\null{eq.~(#1)\null
}

\hypersetup{
    bookmarks=false,         
    unicode=false,          
    pdftoolbar=true,        
    pdfmenubar=true,        
    pdffitwindow=false, 
    pdfstartview={FitH},    
    pdftitle={P fluxes and exotic branes},    
    pdfauthor={Davide M. Lombardo, Fabio Riccioni, Stefano Risoli},     
    pdfsubject={},   
    pdfnewwindow=true,      
    colorlinks=false,       
    linkcolor=blue,          
    citecolor=blue,        
    filecolor=blue,      
    urlcolor=blue           
    linkbordercolor={blue},
    citebordercolor={purple},
    urlbordercolor={gray}
}

\begin{document}

\begin{titlepage}
\begin{center}

\vskip 1.5cm

{\Large \bf $P$ fluxes and exotic branes}

\vskip 1.5cm

{\bf  Davide M. Lombardo\,$^1$, Fabio Riccioni\,$^2$ and Stefano Risoli\,$^{1,2}$}

\vskip 30pt

{\em $^1$ \hskip -.1truecm  Dipartimento di Fisica, Universit\`a di Roma ``La Sapienza'',\\ Piazzale Aldo Moro 2, 00185 Roma, Italy
 \vskip 5pt }

\vskip 15pt

{\em $^2$ \hskip -.1truecm
 INFN Sezione di Roma,   Dipartimento di Fisica, Universit\`a di Roma ``La Sapienza'',\\ Piazzale Aldo Moro 2, 00185 Roma, Italy
 \vskip 25pt }

{email addresses: {\tt Lombardo.1651528@studenti.uniroma1.it},  {\tt Fabio.Riccioni@roma1.infn.it}, {\tt Stefano.Risoli@roma1.infn.it}  \vskip 25pt}

\end{center}

\vskip 0.5cm

\begin{center} {\bf ABSTRACT}\\[3ex]
\end{center}
We consider the ${\cal N}=1$ superpotential generated in type-II orientifold models by non-geometric fluxes. In particular, 
we focus on the family of $P$ fluxes, that are related by T-duality transformations to the S-dual of the $Q$ flux. 
We determine the general rule that transforms a given flux in this family under a single T-duality transformation. This rule allows to derive a complete expression for the superpotential for both the IIA and the IIB theory for the particular case of a $T^6/[\mathbb{Z}_2 \times \mathbb{Z}_2 ]$ orientifold. We then consider how these fluxes modify the generalised Bianchi identities. In particular, we derive a fully consistent set of quadratic constraints coming from the NS-NS Bianchi identities. On the other hand, the $P$ flux Bianchi identities induce tadpoles, and we determine a set of exotic branes that can be consistently included in order to cancel them. This is achieved by determining a universal transformation rule under T-duality satisfied by all the branes in string theory.

\end{titlepage}

\newpage
\setcounter{page}{1} \tableofcontents

\vskip 2truecm

\setcounter{page}{1} \numberwithin{equation}{section}

\section{Introduction}
Fluxes play a crucial role in all phenomenological aspects of string theory, because their presence in general induces a potential for the moduli, which can thus be suitably stabilised \cite{fluxes} (for a review, see {\it e.g.} \cite{fluxreview}). We are interested in orientifold type-II Calabi-Yau compactifications with fluxes turned on, whose low-energy effective actions are ${\cal N}=1$ supergravity theories in four dimensions with a given superpotential determined by the fluxes. In the case of IIB O3-orientifolds, only the  NS-NS and RR 3-form fluxes $H_3$ and $F_3$ can be turned on, and they generate the Gukov-Vafa-Witten superpotential~\cite{Gukov:1999ya}
\begin{equation}
 { W_{\rm IIB/O3}=  \int ( F_3 - i S H_3 ) \wedge \Omega  } \quad , \label{GVWsuperpotential}
\end{equation}
where $\Omega$ is the holomorphic 3-form of the Calabi-Yau manifold and $S$ is the axion-dilaton. For IIA O6-orientifolds, one can in general turn on all RR fluxes from $F_0$ to $F_6$ together with the NS-NS 3-form flux and the metric flux $f_{ab}^c$, and the superpotential reads  \cite{Shelton:2005cf,Aldazabal:2006up,Villadoro}
\begin{equation}
W_{\rm IIA/O6} = \int [ e^{J_{\rm c}} \wedge F_{RR} + \Omega_{\rm c} \wedge (H_3 + f \cdot J_{\rm c} )] \quad ,\label{IIAgeometricsuperpotential}
\end{equation}
where with $J_{\rm c}$ and $\Omega_{\rm c}$ one denotes the complexified K\"{a}hler form and the holomorphic 3-form, and $( f \cdot J_{\rm c} )_{abc} = 3f_{[ab}^d (J_{\rm c})_{c]d}$.

In this paper we will focus on the specific case of a  $T^6/[\mathbb{Z}_2 \times \mathbb{Z}_2 ]$ orientifold, of which we now review the standard notation used in \cite{Aldazabal:2006up} (see also \cite{Shelton:2005cf,Guarino,Aldazabal:2008zza}) in order to make the analysis in the rest of the paper more clear. One factorises the 6-torus as ${T^6 = \bigotimes_{i=1}^3 T_{(i)}^2}$, and the two $\mathbb{Z}_2$'s act as $(-1,-1,1)$ and $(1,-1,-1)$ respectively on the coordinates of the three 2-tori. Denoting these coordinates as ${(x^i , y^i )}$, one defines the three 2-forms ${ \omega_i = -dx^i \wedge d y^i }$ as the natural basis for closed 2-forms, while the basis for closed 4-forms is given by the Hodge duals $\tilde{\omega}_i = * \omega_i$. The K\"{a}hler form $J$ and the holomorphic 3-form $\Omega$ are given by the expressions
\begin{equation}
J = \sum_{i=1}^3 A_i \omega_i \qquad \Omega =(d x^1 + i \tau_1 d y^1 ) \wedge (d x^2 + i \tau_2 d y^2 )\wedge (d x^3 + i \tau_3 d y^3 )  \quad ,
\end{equation}
where $A_i$ and $\tau_i$ are the volumes and complex structure moduli of the three different tori.  The orientifold projection acts like ${\Omega_P (-1)^{F_L} \sigma}$, where $\Omega_P$ is the world-sheet parity reversal, $F_L$ is the world-sheet left-mover fermionic number and $\sigma$ is a space-time involution. 
In the IIB case, the involution acts on the coordinates as
\begin{equation}
\sigma_B (x^i ) = - x^i \qquad \sigma_B ( y^i ) = - y^i \quad ,
\end{equation}
and the untwisted moduli\footnote{One can construct freely acting $\mathbb{Z}_2 \times \mathbb{Z}_2$ orbifolds \cite{freelyacting} such that the twisted sector does not contain massless scalars. Orientifolds of freely-acting orbifolds have  been constructed in {\it e.g.} \cite{Antoniadis:1998ki}. For a review on orientifold models, see {\it e.g.} \cite{Angelantonj:2002ct}.} are the axion-dilaton $S$, the complex-structure moduli $U_i$ that simply coincide with the toroidal complex structures, {\it i.e.} $U_i = \tau_i$, and the complex K\"{a}hler moduli $T_i$ which are given in terms of the  K\"{a}hler form and the RR 4-form by the expression
\begin{equation}
{{\mathcal J}_{\rm c} = C_4 + \frac{i}{2} e^{-\phi} J \wedge J = i \sum_i T_i \tilde{\omega}_i} \quad .
\end{equation} 
In the IIA O6-orientifold, instead, the action of the involution $\sigma_A$ is  
\begin{equation}
\sigma_A (x^i ) =  x^i \qquad \sigma_A ( y^i ) = - y^i \quad .
\end{equation}
This implies that the $\tau_i$'s are real. The real   part of the $S$ and $U_i$ moduli  consists of the $\tau_i$'s and the dilaton, while their imaginary part consists of the RR 3-form $C_3$. The complexified holomorphic 3-form has the expression 
\begin{align}
 \Omega_{\rm c} & =  i S  (dx^1 \wedge dx^2 \wedge dx^3 ) - i U_1 (  dx^1 \wedge dy^2 \wedge dy^3 ) \nonumber \\
&  - i U_2 ( dy^1 \wedge dx^2 \wedge dy^3 ) -  i U_3 ( dy^1 \wedge dy^2 \wedge dx^3 ) \quad ,
\end{align}
which is therefore linear in both $S$ and $U_i$. 
In the IIA case, it is the $B$ field that complexifies the K\"{a}hler form, so that the $T_i$ moduli are given by the expression
\begin{equation}
{ J_{\rm c} = B + i J = i \sum_i T_i {\omega}_i} \quad .
\end{equation}
The two orientifold models are mapped into each other by performing three T-dualities along the $x^i$ directions, under which operation the moduli $U_i$ and $T_i$ are interchanged.  This operation corresponds to mirror symmetry for this specific orbifold~\cite{Strominger:1996it}. 

If one turns on the RR fluxes, it can be easily seen from eqs. \eqref{GVWsuperpotential} and \eqref{IIAgeometricsuperpotential} that one generates 
 a term in the superpotential which is a cubic polynomial in the $U$ moduli from the IIB perspective and in  the $T$ moduli from the IIA perspective. 
The RR fluxes are related by T-duality as   
\begin{equation}
F_{a b_1 ...b_p} \overset{T_a}{\longleftrightarrow} F_{b_1 ...b_p} \quad , \label{TdualityruleRRfluxes}
\end{equation} 
where with $a$ and $b$ we denote any of the internal directions.  In IIB, only the 3-form flux $F_3$ is turned on,  and performing three T-dualities along the $x$ directions,  this is mapped to the various fluxes of IIA according to how many indices there are along the $x$ directions. The result is summarised in Table \ref{TableRRfluxes}.
\begin{table}[t!]
\begin{center}
\begin{tabular}{|c||c|c|}
\hline \rule[-1mm]{0mm}{6mm} RR flux & IIB  & IIA   \\
\hline \hline \rule[-1mm]{0mm}{5mm} 
 $-m$ &  $F_{x^1 x^2 x^3}$ &   ${F}$    \\
  \rule[-2mm]{0mm}{6mm}
$-q_i$ &  $F_{y^i x^j x^k}$ & $F_{x^i y^i}$   \\
  \rule[-2mm]{0mm}{6mm}
$e_i$ & $ F_{x^i y^j y^k}$ & $F_{x^j y^j x^k y^k}$   \\ 
 \rule[-2mm]{0 mm}{6mm}
 $-e_0$ & $F_{y^1 y^2 y^3 }$ & $F_{x^1 y^1 x^2 y^2 x^3 y^3}$ \\ 
\hline 
\end{tabular}
\caption{\footnotesize The RR fluxes in the IIB and IIA setup.  In each row, the two fluxes are related by three T-dualities along the $x$ directions. The indices $i,j,k$ denoting the three tori are always meant to be different, and for the fluxes $F_{y^i x^j x^k}$ and $ F_{x^i y^j y^k}$ the indices are meant to be in the right cyclic order ({\it i.e.} for instance if $i=1$, then $j=2$ and $k=3$). These rules apply to all the other fluxes we will write down in this paper. In the first column we list the names with which we identify the fluxes, following \cite{Aldazabal:2006up}.} \label{TableRRfluxes}
\end{center}
\end{table}
In other words, the IIA and IIB theory give rise to the same model provided that the fluxes of the two theories are identified as in Table \ref{TableRRfluxes}. The fluxes are precisely those that are allowed by the orientifold projection in each of the two theories. 

The situation is different when NS-NS geometric fluxes are turned on. In this case, the $H_3$ flux in \eqref{GVWsuperpotential} gives a term which is $S$ times a cubic polynomial in the $U$ moduli for the IIB theory, while in the IIA theory the $H_3$ flux in \eqref{IIAgeometricsuperpotential} gives a term linear in $S$ and one linear in $U$, while the $f$ flux gives a term linear is $ST$ and one linear in $UT$. Thus obviously the two models cannot be identified by mirror symmetry. This is not surprising, because  indeed these fluxes are related to the non-geometric $Q$ and $R$ fluxes by the chain of T-dualities~\cite{Shelton:2005cf,Aldazabal:2006up}
\begin{equation}
H_{abc} \overset{T_c}{\longleftrightarrow} -f_{ab}^c \overset{T_b}{\longleftrightarrow} -Q_a^{bc} \overset{T_a}{\longleftrightarrow} R^{abc}  \quad . \label{TdualityruleNSfluxes}
\end{equation} 
In particular, in our model this implies that in IIB both the $H$ and $Q$ fluxes can be turned on, and they are related by T-duality to the IIA fluxes as in Table \ref{TableNSfluxes}.
The superpotential for IIB then becomes \cite{Shelton:2005cf,Aldazabal:2006up}~\footnote{The convention for the contraction of indices for the non-geometric fluxes will be given explicitly in section 2.}   
\begin{equation}
 { W_{\rm IIB/O3}=  \int ( F_3 - i S H_3 + Q \cdot {\mathcal J}_{\rm c}) \wedge \Omega  } \quad , \label{NSnongeomIIBsuperpot}
\end{equation}
while in IIA one has
\begin{equation}
W_{\rm IIA/O6} = \int [ e^{J_{\rm c}} \wedge F_{RR} + \Omega_{\rm c} \wedge (H_3 + f \cdot J_{\rm c} + Q \cdot J_{\rm c}^{2} + R \cdot J_{\rm c}^{3} )] \quad . \label{NSnongeomIIAsuperpot}
\end{equation}
This includes now a term proportional to $T$ times a cubic polynomial in $U$ for the IIB theory, and terms of the form $ST^2$, $ST^3$, $UT^2$ and $UT^3$ for the IIA theory. This implies that the two models are dual, where the duality exchanges the $U$ and $T$ moduli and maps the fluxes as in Table \ref{TableNSfluxes}.
\begin{table}[t]
\begin{center}
\begin{tabular}{|c||c|c|}
\hline \rule[-1mm]{0mm}{6mm} NS-NS flux & IIB  & IIA \\
\hline \hline \rule[-0.5mm]{0mm}{5mm} 
$\bar{h}_0$ &  $H_{x^1 x^2 x^3}$ &   ${ R^{x^1 x^2 x^3}}$  \\
  \rule[-3mm]{0mm}{6mm}
$-\bar{a}_i$ &  $H_{y^i x^j x^k}$ & ${-Q_{y^i}^{x^j x^k}}$   \\
  \rule[-3mm]{0mm}{6mm}
$-a_i$ & $ H_{x^i y^j y^k}$ & $- f_{y^j y^k}^{x^i}$  \\
 \rule[-3mm]{0mm}{6mm}
$h_0$&  $H_{y^i y^j y^k}$ & $H_{y^i y^j y^k}$    \\
\hline \rule[-1mm]{0mm}{6mm} 
$-b_{ii}$ & ${Q_{x^i}^{x^j x^k}}$ & $f_{x^j x^k}^{x^i}$   \\
 \rule[-2mm]{0mm}{6mm}
$b_{ij}$  & ${Q_{y^j}^{x^k y^i}}$ & $f_{y^j x^k}^{y^i}$  \\
 \rule[-2mm]{0mm}{6mm}
$-h_i$ & ${Q_{y^i}^{x^j x^k}}$ & $-H_{y^i x^j x^k }$   \\
 \rule[-2mm]{0mm}{6mm}
$-\bar{b}_{ij}$& ${Q_{x^j}^{x^i y^k}}$ & ${ Q_{x^i}^{y^k x^j}}$    \\
 \rule[-2mm]{0mm}{6mm}
$-\bar{h}_i$  & ${Q_{x^i}^{y^j y^k}}$ & ${ R^{ x^i y^k y^j}}$   \\
 \rule[-2.5mm]{0mm}{6mm}
 $-\bar{b}_{ii}$ &  ${Q_{y^i}^{y^j y^k}}$ &  ${ Q_{y^i}^{y^j y^k}}$  \\
\hline
\end{tabular}
\caption{\footnotesize The IIB NS-NS fluxes and their IIA duals. As in \cite{Aldazabal:2006up} we identify the fluxes with the notation given in the first column. The conventions for the indices are explained in the caption of Table \ref{TableRRfluxes}. } \label{TableNSfluxes}
\end{center}
\end{table}

The IIB superpotential of eq. \eqref{GVWsuperpotential} possesses the nice property that it transforms correctly under S-duality. This property is not shared by the superpotential in eq. \eqref{NSnongeomIIBsuperpot} where the $Q$ flux is included. In \cite{Aldazabal:2006up} it was shown that one has to add the flux $P_a^{bc}$, which is the S-dual of the $Q$ flux.
This leads to the superpotential
\begin{equation}
 { W_{\rm IIB/O3}=  \int ( [F_3 - i S H_3] + [(Q - iSP )\cdot {\mathcal J}_{\rm c}]_3 ) \wedge \Omega  } \quad ,\label{IIBsuperpotentialwithonePflux}
 \end{equation}
and the additional term $S P \cdot {\mathcal J}_{\rm c} \wedge \Omega$ generates a term linear in $ST$ times a cubic polynomial in $U$.
Although adding this $P$ flux allows one to recover S-duality invariance, the duality with the IIA model is spoiled. 
In \cite{Aldazabal:2010ef} the most general form for the superpotential of the IIB theory was written down using generalised geometry techniques\footnote{For a discussion on the superpotential in generalised geometry see  \cite{Grana:2006hr}.} in a general case of a Calabi-Yau O3-orientifold.
This superpotential includes naturally fluxes that  are related to the $P_a^{bc}$ flux by perturbative duality transformations. In terms of the $S$, $T$ and $U$ moduli, this gives a general expression which is at most linear in $S$ while it is a cubic polynomial in both $T$ and $U$. Anyway, an explicit T-duality transformation for the $P$ flux, analogous to the ones in eqs. \eqref{TdualityruleRRfluxes} and \eqref{TdualityruleNSfluxes}, was not given, and therefore the IIA-equivalent expression for the superpotential was not derived.  

One of the aims of this paper is precisely to perform this additional step. We will first determine how in general the $P_a^{bc}$ flux transforms under T-duality,  making use of the results of \cite{Bergshoeff:2015cba}, where the complete family of $P$ fluxes was determined in any dimensions.
We will then apply these new T-duality rules to build the terms of the superpotential that include all the allowed  $P$ fluxes both in IIB and IIA theories. This produces an explicit polynomial form for the superpotential with an additional term proportional to $T^2$ times a cubic polynomial in $U$ in the IIB theory, while in IIA one generates a term proportional to $SU$ and one proportional to $U^2$, both times cubic polynomials in $T$. As a result, the expression for the resulting superpotential of each theory is mapped to the one of the other theory provided that the moduli $U$ and $T$ are interchanged and the $P$ fluxes in the two theories are identified according to our rule, which amounts to the statement that the 
duality between the IIB and IIA model is restored.  The IIB expression for the superpotential  coincides with the expression of \cite{Aldazabal:2010ef} as far as the $P$ fluxes are concerned. 

Turning on fluxes in general modifies the Bianchi identities for the  field strengths of the various potentials in the theory in such a way to produce an effective charge for these potentials which is given by the integral in the internal directions of a quadratic term in the fluxes. In particular, the absence of NS branes leads to a set of quadratic constraints for all the NS-NS fluxes~\cite{Shelton:2005cf,Aldazabal:2006up,wrase,penas,shukla1,andriot,shukla2}. By using our T-duality rules, we will manage to show how these constraints are modified by the inclusion of terms containing both RR and $P$ fluxes. The RR Bianchi identities, instead, can give a non-vanishing contribution to the effective number of D-branes, which has to be taken into account in the tadpole cancellation condition. In IIB, turning on the $P_a^{bc}$ flux leads to a new tadpole that is cancelled by the addition of a proper number of  7-branes which are the S-duals of the D7-branes~\cite{Aldazabal:2006up}.  By turning on more general $P$ fluxes, one then expects that additional branes, which are related by suitable T-duality transformations to the S-dual of the D7-brane, can be included. 

In a series of papers \cite{Bergshoeff:2010xc,stringsolitons,axel,dominantweights,bergshoeffriccionimarrani}, all the 1/2-BPS branes that are present in string theory in any dimension have been classified according to their properties with respect to the duality symmetry group of the theory. By carefully analysing this classification, we will manage to determine a 
 universal T-duality transformation rule for all the branes in string theory. This  will be the second main result of this paper. This rule will allow us to write down all the branes that are sourced by the $P$ fluxes, and thus all possible tadpole conditions that have to be imposed. 
In general, by duality a brane in a given dimension can be mapped to a so called {\it exotic} brane, which is an object that in  the higher-dimensional theory  is  a generalised KK-monopole, {\it i.e.} an object well-defined only in the presence of isometries \cite{Elitzur:1997zn,LozanoTellechea:2000mc,deboer,zagermann,sakatani}. This is precisely the case for all the branes that are related by T-duality to the S-dual of the D7-brane, and as a consequence the tadpole conditions that we find require in general the inclusion of exotic branes.

The paper is organised as follows.  In section 2 we will first review the results of \cite{Bergshoeff:2015cba} by writing down all the $P$ fluxes in both the IIB and IIA theories, and  then we  will derive a universal T-duality rule for such fluxes which will allow us to write down the most general form for the superpotential in both the IIB and IIA theories. In section 3 we will derive a universal T-duality rule for all the branes in string theory, and we will determine all the exotic branes that can be included in the orientifold model together to the S-dual of the D7-brane.
In section 4 we will apply our T-duality rules to determine how the $P$ fluxes modify the NS-NS Bianchi identities, and then to derive how they give rise to new tadpole conditions for the exotic branes discussed in section 3. Finally, section 5 contains our conclusions.

\section{$P$ fluxes and the superpotential}

The aim of this section is to derive the expression of the superpotential containing all the allowed $P$ fluxes for the IIB and IIA orientifold models discussed in the introduction. We will first review the results of \cite{Bergshoeff:2015cba}, where the complete family of $P$ fluxes was derived in all dimensions. We will then derive the rule to transform each of these fluxes under T-duality, and finally we will use this result to write down the superpotential. In  \cite{Aldazabal:2010ef} (see also \cite{Aldazabal:2008zza}) the most general form of the superpotential generated by geometric and non-geometric fluxes in IIB was derived. We will show that the T-duality rules that we find reproduce the same $P$ flux contribution for IIB, and we will also show how the same superpotential can be written in terms of the IIA fluxes. The inclusion of $P$ fluxes has been considered in the literature in various contexts, see {\it e.g.} \cite{Pfluxes}.

From the point of view of the four-dimensional effective action, fluxes give rise to gaugings, and the way in which a particular flux transforms  by the action of the duality symmetry of the ungauged theory is encoded in the so-called `embedding tensor' \cite{Nicolai:2000sc}.  We will consider the embedding tensor of the maximal supergravity theory, and then we will  take into account only the components  that survive in the ${\cal N} =1$ model. The representations of the global symmetry group of any maximal supergravity theory to which the embedding tensor belongs were derived in a series of papers 
\cite{embeddingtensorreprs}. In particular, in four dimensions the embedding tensor belongs to the ${\bf 912}$ of $E_{7(7)}$. One is then interested in decomposing this representation in terms of the perturbative symmetry $SO(6,6)$ of the global symmetry group. Considering the embedding $E_{7(7)} \supset SO(6,6) \times SL(2,\mathbb{R})$, this representation decomposes as
\begin{equation}
{\bf 912} = {\bf  (32,3)} \oplus {\bf (220,2)} \oplus {\bf (12,2)} \oplus   {\bf (352,1)} \quad . 
\label{912decomposition}
\end{equation}
The $SL(2,\mathbb{R})$ symmetry is the one that transforms non-linearly the complex scalar made of the four-dimensional dilaton and the axion dual to the NS-NS 2-form. By further decomposing $SL(2,\mathbb{R}) \supset \mathbb{R}^+$, one therefore associates the $\mathbb{R}^+$ weight with the dilaton weight. In particular one has
\begin{align}
& {\bf  (32,3)} = {\bf  32_2} \oplus {\bf  32_0} \oplus {\bf  32_{-2}} \nonumber \\
& {\bf  (220,2)} = {\bf  220_1} \oplus {\bf  220_{-1}} \\
& {\bf  (12,2)} = {\bf  12_1} \oplus {\bf  12_{-1}} \quad ,\nonumber
\end{align}
where the subscript denotes the $\mathbb{R}^+$ weight.
The representation ${\bf  32_2}$ corresponds to the RR fluxes, the ${\bf  220_1}$ corresponds to the NS-NS fluxes and the  $ {\bf  352_0}$ corresponds to the $P$ fluxes. This can be seen by decomposing each of the $SO(6,6)$ representations  in terms of $GL(6,\mathbb{R})$, which we do in detail now. 

It is straightforward to see how this decomposition works for the case of the RR fluxes. The ${\bf 32}$ is the spinorial representation $\theta_\alpha$ of $SO(6,6)$,  and according to the convention that one chooses for the chirality of this spinor, one has the two possible decompositions
\begin{equation}
\theta_\alpha  \rightarrow \left\{ \begin{array}{ll} F_a \ \  F_{abc} \ \  F_{abcde} \ \  \ \ \  \ \ \ \ \ \ ({\rm IIB})
        \\
        \\
         F \ \  F_{ab} \ \  F_{abcd}  \ \ F_{abcdef} \ \ \ ({\rm IIA})\end{array} \right.   \quad , \label{decompositionofRRfluxes}
\end{equation}
corresponding to the RR fluxes of odd rank in the IIB theory and of even rank in the IIA theory. Obviously this representation contains only  geometric fluxes. The T-duality rule given in eq. \eqref{TdualityruleRRfluxes} maps a given flux in one theory to a flux in the other theory. All the components in eq. \eqref{decompositionofRRfluxes} are connected by chains of T-duality transformations.

The next representation is the ${{\bf 220 }}$, which is the representation $\theta_{MNP}$ of $SO(6,6)$ with three antisymmetrised vector indices. This corresponds to the NS-NS fluxes, and indeed the embedding tensor decomposes under $GL(6,\mathbb{R})$ as 
\begin{equation}
\theta_{MNP} \ \rightarrow \ H_{abc} \quad f_{ab}^c \quad Q_{a}^{bc} \quad R^{abc} \quad .\label{decompositionofNSfluxes}
\end{equation}
The T-duality rule given in eq. \eqref{TdualityruleNSfluxes} connects the different components in the equation above, but in this case not all the components can be reached by chains of T-dualities starting for instance from a given  $H_3$ flux. More precisely, the components that are not connected by T-duality to a given $H_3$ flux are $f_{ab}^a$ and $Q_a^{ab}$ (with indices not summed). It is common procedure in the literature not to consider these fluxes, and we  will  also not consider them in this paper.

We then move to the representation of the $P$ fluxes, which is the ${ \bf 352}$ of ${ SO(6,6)}$.
This is  the vector-spinor ({\it i.e.} `gravitino') representation  ${\theta_{M \dot{\alpha}}}$. 
By decomposing the whole representation under ${GL(6,\mathbb{R})}$ one gets~\cite{Bergshoeff:2015cba}
\begin{equation}
\theta_{M \dot{\alpha}}  \rightarrow \left\{ \begin{array}{ll}  P_a \ \ P_a^{b_1 b_2}\  \ P_a^{b_1 ...b_4}\  \ P_a^{b_1 ...b_6}\   \ P^{a,b_1 b_2} \ \ P^{a, b_1 ...b_4} \ \  P^{a, b_1 ...b_6}  \qquad ({\rm IIB})
        \\
        \\
        P_a^b \  \ P_a^{b_1 b_2 b_3} \ \ P_a^{b_1 ...b_5} \ \ P^{a,b} \ \ P^{a,b_1 b_2 b_3} \ \ P^{a, b_1 ...b_5} \qquad \qquad \quad \ \ \! \ ({\rm IIA})
       \end{array} \right.   \quad , \label{decompositionofPfluxes}
\end{equation}
where the convention for each of the two decompositions is fixed  by the corresponding convention of the spinor, which is given in eq. \eqref{decompositionofRRfluxes}. The flux  $P_a^{b_1 b_2}$, which is the second flux in the IIB decomposition, is the S-dual of the $Q$ flux. In all the fluxes, the indices $b_1...b_p$ are completely antisymmetrised, and the representations with all upstairs indices $a , b_1 ...b_p$ are irreducible with vanishing completely antisymmetric part, while the representations with the $a$ index downstairs and some $b$ indices upstairs are reducible, with the condition that the singlet is always removed~\cite{Bergshoeff:2015cba}.

To conjecture how a single T-duality transformation should act on a given $P$ flux within the set of fluxes in eq. \eqref{decompositionofPfluxes}, we simply observe that  in the embedding tensor $\theta_{M\dot{\alpha}}$ the vector index transforms as it should, namely a lower index in a given direction is raised if one performs a T-duality in that direction, while the spinor index decomposes in the set of all even  or odd antisymmetric indices, and T-duality should remove or add an index according to whether it is already there or not. 
As a consequence, one derives the following T-duality rules
\begin{align}
P_a^{b_1 ... b_p} \ &  \overset{T_a}{\longleftrightarrow} \ P^{a, b_1 ... b_p a} \nonumber \\
P_a^{b_1 ... b_p} \ &  \overset{T_{b_p}}{\longleftrightarrow} \ P_{a}^{b_1 ... b_{p-1}}  \label{TdualityrulesPfluxes}\\
P^{a ,b_1 ... b_p} \  &  \overset{T_{b_p}}{\longleftrightarrow} \ P^{a , b_1 ... b_{p-1}} \quad ,\nonumber 
\end{align}
which simply summarise the statements above, that is under the action of $T_a$ a downstairs $a$ index is raised and vice versa, while in the set of antisymmetric indices the rule is precisely as for the RR fluxes.

The components of the flux $P_a^{b_1 b_2}$ that one considers are such that $b_1$ and $b_2$ are different from $a$, precisely as for the $Q$ flux. Therefore, by applying the T-duality rules in eq. \eqref{TdualityrulesPfluxes}, one finds that by performing any chain of T-dualities one  always ends up with components such that if the ${a}$ index is down, then it is different from any of the ${ b}$ indices, while if it is up it has to be parallel to the ${ b}$ indices.  It is for this reason that in eq. \eqref{TdualityrulesPfluxes} we have not included the rule that maps the flux $P_a^{b_1 ... b_p a}$ to $P^{a, b_1 ...b_p}$ under $T_a$: both these components are not connected by T-duality transformations to the components of the flux $P_a^{b_1 b_2}$ we are considering.\footnote{For the same reason, the flux $P_a^{b_1 ...b_6}$ of the IIB theory will not be considered in our four-dimensional model because in this case the index $a$ cannot be different from all the indices $b$.}
It should also be appreciated that all these rules actually apply to any dimension, although in this paper we are only interested in the four-dimensional case. 

We can now apply these rules to determine all the $P$ fluxes that can be included in the four-dimensional $T^6/[\mathbb{Z}_2 \times \mathbb{Z}_2 ]$ orientifold model. In the case of the O3-orientifold of IIB, as we have already reviewed in the introduction all the fluxes $P_a^{bc}$ with each of the three indices along a direction of each of the three different tori can be turned on \cite{Aldazabal:2006up}.
 Using the T-duality rules of eq. \eqref{TdualityrulesPfluxes}, we can then apply three T-duality transformations along the three $x$ directions to determine the corresponding fluxes from the IIA perspective. The result is listed in the upper half of Table \ref{allPfluxes}. As can be seen from the table, in the IIA picture one turns on some components of all the $P$ fluxes of the IIA theory that are listed in eq. \eqref{decompositionofPfluxes}.

\begin{table}[t!]
\begin{center}
\begin{tabular}{|c||c|c|}
\hline \rule[-1mm]{0mm}{6mm} $P$ flux & IIB & IIA   \\
\hline \hline 
 \rule[-1mm]{0mm}{6mm} 
$f_i$ & $P_{y^i}^{x^k x^j}$ & $P_{y^i}^{x^i}$   \\
 \rule[-2mm]{0mm}{6mm} 
$g_{ji}$  & $P_{y^i}^{y^j x^k}$ & $P_{y^i}^{x^i x^j y^j} $   \\
 \rule[-2mm]{0mm}{6mm} $-\bar{g}_{ii}$ & $P_{y^i}^{y^j y^k}$   & $P_{y^i}^{x^i x^j x^k y^j y^k}$   \\
\rule[-2mm]{0mm}{6mm} $g_{ii}$ & $P_{x^i}^{x^k x^j}$ &$P^{x^i , x^i}$ \\
 \rule[-2mm]{0mm}{6mm} 
$\bar{g}_{ki}$ & $P_{x^i}^{y^j x^k}$ & $P^{x^i, x^i x^j y^j}$    \\
 \rule[-2mm]{0mm}{6mm} $-\bar{f}_i$ & $P_{x^i}^{y^j y^k}$  & $P^{x^i , x^i x^j x^k y^j y^k}$   \\ 
\hline \rule[-1mm]{0mm}{6mm}  $-f_i'$ & $P^{x^i, x^i x^j x^k y^i}$     & $P_{x^i}^{y^i}$  \\
 \rule[-1mm]{0mm}{6mm} $g'_{ki}$  & $P^{x^i, x^i x^j y^i y^k}$     & $P_{x^i}^{y^i x^k y^k}$   \\
 \rule[-1mm]{0mm}{6mm} $-\bar{g}'_{ii}$  &  $P^{x^i, x^i y^i y^j y^k}$     & $P_{x^i}^{y^i x^j y^j x^k y^k}$  \\
\rule[-1mm]{0mm}{6mm}  $-g'_{ii}$ & $P^{y^i, x^i x^j x^k y^i}$     & $P^{y^i,  y^i}$   \\
  \rule[-1mm]{0mm}{6mm} $\bar{g}'_{ki}$  &  $P^{y^i, y^i x^i y^j x^k}$     & $P^{y^i, y^i x^j y^j}$  \\
 \rule[-1mm]{0mm}{6mm} $\bar{f}'_i$ & $P^{y^i, y^i y^j y^k x^i}$     & $P^{y^i, y^i x^j x^k y^j  y^k}$  \\
\hline
\end{tabular}
\caption{\footnotesize Table containing all the $P$ fluxes that can be turned on in the ${\cal N}=1$ orientifold model for both IIB and IIA. The convention for the indices is explained in the caption of Table \ref{TableRRfluxes}. In the first column we list the notation that we  use to identify each of the fluxes in both theories.}
\label{allPfluxes}
\end{center}
\end{table}

It is straightforward to deduce the rule for the $P$ fluxes that survive the orientifold projection in the IIA theory. With respect to $\Omega_P (-1)^{F_L}$, the fluxes $P^{a,b}$, $P_a^{b_1 b_2 b_3}$ and $P^{a,b_1 ...b_5}$ are even, while the fluxes $P_a^b$, $P^{a,b_1 b_2 b_3}$ and $P_a^{b_1 ...b_5}$ are odd. For the fluxes that are even, one must have an even number of $x$ and an even number of $y$ indices, while for the fluxes that are odd both the number of $x$ and $y$ indices have to be odd. The indices of the fluxes $P_a^{b_1 ....b_p}$ are always grouped in pairs of indices belonging to a given torus, while for the fluxes $P^{a,b_1 ...b_p}$ there is always one index (the $a$ index) which is always an $x$ index and it is  repeated, while again the other indices are grouped in pairs belonging to a given torus. 

The $a$ index in the fluxes $P_a^{b_1 ...b_p}$ in the IIA upper half of Table \ref{allPfluxes} is in all cases a $y$ index, but by analogy with Table \ref{TableNSfluxes} we assume that it can also be $x$, provided that again all indices are grouped in pairs as before. This leads to   
the first three fluxes in the lower half of Table \ref{allPfluxes}. Now, by performing again three T-dualities along the $x$ directions and using the T-duality rules of eq. \eqref{TdualityrulesPfluxes} we see that in the IIB theory these fluxes are mapped to $P^{a, b_1 ...b_4}$. In particular, the $a$ index is $x$ and it is repeated, while the other three indices are each on a different torus. Given that in the IIB theory the orientifold projection acts in the same way on the $x$ and $y$ indices, we include the last three fluxes in the IIB side of the lower half of Table \ref{allPfluxes}, which are mapped in the IIA setting to the fluxes $P^{a,b_1 ...b_p}$ where now the repeated index is a $y$ index and all the others  are again grouped pairwise as before.

We can now  write down the superpotential  which contains all the $P$ fluxes compatible with the orientifold for both the IIB and IIA theory.
We start by considering the IIB superpotential with the $P_a^{bc}$ flux turned on, which is given in eq. \eqref{IIBsuperpotentialwithonePflux}. Performing three T-dualities along the $x$ directions and using the  T-duality rules given in eq. \eqref{TdualityrulesPfluxes} we find a T-dual IIA expression which contains all the IIA $P$ fluxes in the upper half of Table \ref{allPfluxes}.  We then extend this to include also the IIA fluxes in the lower half of the table. Collecting all the terms in a compact notation, we arrive at the expression
\begin{align}
 W_{\rm IIA/O6} & = \int e^{J_{\rm c}} \wedge  F_{RR} + \Omega_{\rm c} \wedge   \Big( H_3 + f \cdot J_{\rm c} +Q \cdot J_{\rm c}^2 + R \cdot J_{\rm c}^3 - P_1^1 \cdot \Omega_{\rm c}  \nonumber \\
& + (P^{1,1} - P_1^3 ) \cdot \Omega_{\rm c} \cdot J_{\rm c} -(P^{1,3} + P_1^5 ) \cdot \Omega_{\rm c} \cdot J_{\rm c}^2 - P^{1,5} \cdot \Omega_{\rm c} \cdot J_{\rm c}^3 \Big) \quad .  
\label{Waallpfluxes}
\end{align}
In this equation, the contraction rules for the NS fluxes are~\cite{Shelton:2005cf,Aldazabal:2006up}
\begin{align}
& ( f \cdot J_{\rm c} )_{abc} = 3f_{[ab}^d (J_{\rm c})_{c]d}\nonumber \\
& ( Q \cdot J_{\rm c}^2 )_{abc} = \tfrac{3}{2}   Q^{de}_{[a}(J_{\rm c}^2)_{bc]de}\\
&  (R  \cdot J_{\rm c}^3)_{abc}= \tfrac{1}{6} R^{def}(J_{\rm c}^3)_{abcdef}  \quad ,\nonumber 
\end{align}
while the rules for the $P$ fluxes are
\begin{align}
& (P_1^1\cdot\Omega_{\rm c})_{abc}  =  \tfrac{3}{2} P^d_{[a}(\Omega_{\rm c})_{bc]d} \nonumber \\
& (P^{1,1}\cdot\Omega_{\rm c} \cdot J_{\rm c})_{abc}   = \tfrac{3}{2}P^{d,d}(\Omega_{\rm c})_{d[ab}(J_{\rm c})_{c]d}\nonumber \\
& (P_1^3\cdot\Omega_{\rm c} \cdot J_{\rm c})_{abc} = \tfrac{1}{2}\Big( \tfrac{3}{4} P^{d_1 d_2 d_3}_{[a}(\Omega_{\rm c})_{bc]d_1}(J_{\rm c})_{d_2 d_3}- \tfrac{3}{2} P^{d_1 d_2 d_3}_{[a}(\Omega_{\rm c})_{|d_1 d_2|b}(J_{\rm c})_{c]d_3} \Big) \nonumber \\
& (P^{1,3}\cdot\Omega_{\rm c}\cdot J_{\rm c}^2 )_{abc}  = \tfrac{1}{2}\Big( \tfrac{3}{2} P^{d,def}(\Omega_{\rm c})_{de[a}(J_{\rm c}^2)_{bc]df} -\tfrac{3}{4}P^{d,def}(\Omega_{\rm c})_{d[ab}(J_{\rm c}^2)_{c]def} \Big) \nonumber \\
&  (P_1^5\cdot\Omega_{\rm c}\cdot J_{\rm c}^2)_{abc}  =\tfrac{1}{3} \Big( \tfrac{1}{16} P^{d_1...d_5}_{[a}(\Omega_{\rm c})_{bc]d_1}(J_{\rm c}^2)_{d_2...d_5}+\tfrac{1}{8}P^{d_1...d_5}_{[a}(\Omega_{\rm c})_{| d_1 d_2 d_3}(J_{\rm c}^2)_{d_4 d_5|bc]}\nonumber \\ 
& \qquad \qquad  - \tfrac{1}{4}P^{d_1...d_5}_{[a}(\Omega_{\rm c})_{|d_1 d_2|b}(J_{\rm c}^2)_{c]d_3 d_4 d_5} \Big)\nonumber \\
& (P^{1,5}\cdot\Omega_{\rm c}\cdot J_{\rm c}^3)_{abc}  =\tfrac{1}{3} \Big(\tfrac{1}{8}P^{d,d e_1...e_4}(\Omega_{\rm c})_{d e_1 e_2}(J_{\rm c}^3)_{d e_3 e_4 abc}   +\tfrac{1}{2}P^{d,d e_1...e_4}(\Omega_{\rm c})_{d e_1[a}(J_{\rm c}^3)_{bc]d e_2 e_3 e_4}\nonumber \\
&\qquad \qquad -\tfrac{1}{16}P^{d,d e_1...e_4}(\Omega_{\rm c})_{d[ab}(J_{\rm c}^3)_{c]d e_1 ... e_4}\Big) \ .
\end{align}
Performing again three T-dualities along the $x$ directions on eq. \eqref{Waallpfluxes}, we come back to the IIB superpotential with all $P$ fluxes allowed by the projections. Namely, we find
\begin{equation}
W_{\rm IIB/O3} = \int [( F_3 - i S H_3  )  + ( Q - i S P_1^2 ) \cdot \mathcal{J}_{\rm c} - P^{1,4} \cdot \mathcal{J}_{\rm c}^2 ]\wedge \Omega \quad ,
\label{Wballpfluxes} 
\end{equation}
where the contractions for $Q$ and $P_1^2$ are~~\cite{Shelton:2005cf,Aldazabal:2006up}
\begin{align}
& (Q \cdot \mathcal{J}_{\rm c})_{abc}=\tfrac{3}{2} Q^{de}_{[a}(\mathcal{J}_{\rm c})_{bc]de}  \nonumber \\
&  (P_1^2 \cdot  \mathcal{J}_{\rm c})_{abc}=\tfrac{3}{2} P^{de}_{[a}(\mathcal{J}_{\rm c})_{bc]de}\quad ,
\end{align}
while the last contraction is defined as
\begin{equation}
(P^{1,4}\cdot{\mathcal{J}_{\rm c}^2})_{abc} = \tfrac{3}{4}P^{d,d e_1 e_2 e_3}(\mathcal{J}_{\rm c})_{d e_1[ab}(\mathcal{J}_{\rm c})_{c]d e_2 e_3} \quad ,
\end{equation}
which is the only non-vanishing contraction we can introduce consistently with the orientifold. 

The IIB/O3  superpotential  we find  coincides with the one derived in \cite{Aldazabal:2010ef} on the basis of generalised geometry considerations,  and this is a positive test in favour of our   T-duality rules for $P$ fluxes. As far as the IIA/O6 result is concerned, our expression is valid for the specific orbifold model we are considering, but it would be interesting to investigate whether the same superpotential can be derived for more general orientifolds and whether it has a natural explanation in the context of generalised geometry. 

We now want to derive the explicit form of the superpotential in terms of the $S,T,U$ moduli. We use for the fluxes the conventions defined in the first column of Tables \ref{TableRRfluxes}, \ref{TableNSfluxes} and \ref{allPfluxes}. Following \cite{Aldazabal:2006up}, we consider the simpler case of three equivalent tori, {\it i.e.} the isotropic case. 
This greatly simplifies the explicit expressions for the superpotentials in both the IIB and IIA theory. We consistently remove the indices from  all the  fluxes $q_i$, $e_i$, $a_i$, $b_{ij}$, $h_i$, $f_i$, $g_{ij}$ and from the corresponding primed and/or barred fluxes in Tables \ref{TableRRfluxes}, \ref{TableNSfluxes} and \ref{allPfluxes}. We rename $b_{ii}$ and $g_{ii}$ as $\beta$ and $\gamma$, and similarly for the equivalent primed and/or barred fluxes~\cite{Aldazabal:2006up}. 
The IIB/O3 superpotential in eq. \eqref{Wballpfluxes} leads to
\begin{align}
W_{\rm IIB/O3} & = e_0 +3ieU -3qU^2 +imU^3\nonumber \\
& +  S\big( ih_0 -3aU +3i\bar{a}U^2                                                                                                - \bar{h}_0 U^3 \big) \nonumber \\
& +3T \Big( -i h - (2b+\beta ) U + i  (2 \bar{b} + \bar{\beta} ) U^2 + \bar{h} U^3 \Big) \nonumber \\
              & + 3ST \big( -f +i(2g +\gamma)U +  (2\bar{g} + \bar{\gamma} )U^2 -i\bar{f}U^3 \big)  \nonumber \\ 
                & - 3T^2\big( f'   +i(2g'+ \gamma' )U - (2\bar{g}' + \bar{\gamma}' )U^2 -i\bar{f}' U^3 \big) \quad , \label{isotropicsuperpotential}
\end{align}
and it can be shown that the IIA/O6 superpotential in eq. \eqref{Waallpfluxes} gives the same expression with $U$ and $T$ interchanged.

\section{T-duality rules and exotic branes}

In the IIB $T^6/[\mathbb{Z}_2\times \mathbb{Z}_2 ]$ O3 orientifold setup the only non-vanishing NS-NS and RR fluxes are $H_3$, $Q$ and $F_3$. These fluxes induce RR tadpoles, which have to be cancelled by the addition of D-branes. In particular, the D-branes that survive the projections are D3-branes and O3-planes, together with D7-branes, which wrap two of the three tori. The D3 and D7-brane charges are induced by the (generalised) Chern-Simons terms
\begin{equation}
\int C_4 \wedge H_3 \wedge F_3 
\qquad \quad
 \int C_8 \wedge Q \cdot F_3 \quad  , \label{RRD7tadpole}
\end{equation}
where $C_4$ and $C_8$ are the 4-form and 8-form RR potentials.
In the equivalent IIA/O6 description only D6-branes spanning either the three $x$ directions or one $x$ and two $y$ directions, each on a different torus, survive the projections. In this case the charge induced by the flux results from the generalised Chern-Simons term 
\begin{equation} \int C_7 \wedge (- H_3 F_0+ f \cdot F_2 -Q \cdot F_4+R\cdot F_6) \quad . 
\end{equation}
The IIB and IIA tadpole conditions are mapped into each other by performing three T-dualities along the $x$ directions. In particular the D3/O3 tadpole condition is mapped to the condition arising from D6/O6 wrapping the three $x$ directions, while the three D7-brane conditions are mapped to the ones of the other three D6 branes~\cite{Aldazabal:2006up}.

In the IIB theory, one can study how each of the two tadpole conditions transform under S-duality. While the D3 tadpole condition is S-duality invariant,  the D7 brane is mapped to its S-dual, whose tadpole is generated by~\cite{Aldazabal:2006up}
\begin{equation}
\int E_8 \wedge P_1^2 \cdot H_3 \quad ,\label{SdualD7tadpole}
\end{equation}
where  the potential $E_8$ is the S-dual of $C_8$ \cite{Meessen:1998qm}. Our aim is to determine the IIA equivalent of this expression. In order to do this, in this section we will  determine how this 7-brane transforms under T-duality. In a series of papers, the full classification of 1/2-BPS branes in type-II string theory compactified on a torus was achieved by analysing the supersymmetry of the corresponding brane effective action   \cite{Bergshoeff:2010xc,stringsolitons,bergshoeffriccionimarrani}, while the connection of this analysis with specific group theory properties of such branes was shown in \cite{axel,dominantweights}. This classification was also extended to theories with less supersymmetry in \cite{heteroticwrappingrules,Bergshoeff:2012jb}. In the following, we will first review the basic results of this classification and we will then derive a universal T-duality transformation rule  for any brane in string theory, which will give in particular the IIA branes that are related by duality to the S-dual of the D7-brane in our model, together with additional branes in IIB that must be included for consistency. In the next section, these rules will then be used to determine the full set of tadpole conditions for these branes.

Following \cite{Bergshoeff:2010xc}, we introduce the non-positive integer number $\alpha$ occurring in the expression $T \sim g_S^{\alpha}$ that says how the tension of a given brane scales with respect to the string coupling $g_S$. For example, in ten dimensions one has the fundamental string with $\alpha=0$, the D-branes with $\alpha=-1$, the NS5-brane with $\alpha=-2$ and in the IIB theory one also has the S-dual of the D7-brane which has 
$\alpha=-3$. By studying the branes with different values of $\alpha$  in lower dimensions, what one finds is that apart from the branes that arise as dimensional reductions of the branes in ten dimensions, there are additional branes that have no higher dimensional origin. An example of a brane of this type is the $\alpha=-2$ brane that is obtained by performing two T-dualities in two directions transverse to the NS5-brane~\cite{deboer}.  The corresponding solution describes a metric that in these directions in not globally defined, {\it i.e.} a T-fold~\cite{Tfold}, and as such it can  be considered as a well-defined geometric solution only in eight dimensions. In \cite{LozanoTellechea:2000mc}, the $D$-dimensional brane solutions with two transverse directions (so-called  defect branes) obtained by performing U-dualities on brane solutions that arise from ten dimensions have been classified. Typically, from the point of view of the ten-dimensional theory these solutions are locally well-defined only in the presence of isometries, and are dubbed `exotic branes' in the literature~\cite{Elitzur:1997zn}. Denoting with $p+1$ the world-volume directions and with $n$ the number of isometries, such branes are commonly dubbed $p_{-\alpha}^n$-branes. So the $\alpha=-2$ brane with two isometries of \cite{deboer} is a $5^2_2$-brane.     

In the brane classification of \cite{Bergshoeff:2010xc,stringsolitons,bergshoeffriccionimarrani}, the non-geometric nature of a particular brane corresponds to the fact that the brane is charged with respect to a mixed-symmetry potential.\footnote{The mixed-symmetry potentials of the ten-dimensional theories are determined from the infinite-dimensional Kac-Moody algebra $E_{11}$ \cite{WestE11}. In \cite{E11origin} it was shown that by dimensionally reducing such potentials one obtains the full spectrum of forms compatible with the gauge and supersymmetry algebras of the maximal theory  in any dimension.}
In particular, denoting with $A_{p,q,r,..}$ a ten-dimensional mixed-symmetry potential in a representation such that $p,q,r, ...$ (with $p\ \geq q \geq r ...$) denote the length of each column of its Young Tableau, this corresponds to a brane if some of the  indices $p$  are isometries and contain all the indices $q$, which themselves contain all the indices $r$ and so on. In particular, the exotic defect branes discussed in \cite{LozanoTellechea:2000mc} correspond to mixed-symmetry potentials with $p=8$ \cite{defectbranes}, but the analysis of \cite{Bergshoeff:2010xc,stringsolitons,bergshoeffriccionimarrani} is more general because it includes domain walls and space-filling branes by also including mixed-symmetry potentials with $p=9$ and $p=10$. 

The D-branes, {\it i.e.} the $\alpha=-1$ branes, are special because they always arise from D-branes of the ten-dimensional theory. This means that the corresponding potentials are forms, which are indeed the RR forms of the ten-dimensional theory and their duals, together with the 10-form $C_{10}$ associated to the D9-brane in IIB and the 9-form $C_9$ associated to the D8-brane in IIA. In total one gets
\begin{align}
& C_2 \quad C_4 \quad C_6 \quad C_8 \quad C_{10} \quad ({\rm IIB}) \nonumber \\
& C_1 \quad C_3 \quad C_5 \quad C_7 \quad C_9 \quad  \ ({\rm IIA}) \quad . \label{allRRfields}
\end{align}
A single T-duality along the $a$ direction acts on the RR fields in the obvious way: if an $n$-form potential $C_n$ of one theory has no index parallel to $a$, it is mapped to the $(n+1)$-form $C_{n+1}$ of the other theory where the extra index is $a$, and vice versa. It is clear that one obtains all components of all the RR fields listed in eq. \eqref{allRRfields} starting from any particular field and performing repeated T-dualities in all directions.  

The $\alpha=-2$ branes arise from the ten-dimensional mixed-symmetry potentials~\cite{stringsolitons,dualdoubledgeometry}
\begin{equation}
{D_6 \ \ D_{7,1} \ \ D_{8,2} \ \ D_{9,3} \ \ D_{10,4} } \quad , \label{allDpotentials}
\end{equation}
which are the same for the IIB and the IIA theory.
The potential $D_6$ obviously corresponds to the NS5-brane, while the mixed-symmetry potential $D_{7,1}$ corresponds to the KK-monopole and the potential $D_{8,2}$ corresponds to the T-fold of \cite{deboer}. These latter two are solutions with one and two isometric directions respectively, and denoting with $a$ and $b$ these directions one obtains that the potentials corresponding to these solutions are the 6-forms $D_{6 \,  a ,a}$ and $D_{6 \, ab, ab}$. The KK-monopole  solution is obtained by performing a T-duality  on the NS5-brane solution along the direction $a$ transverse to the brane, while the T-fold solution is obtained by T-dualising the KK-monopole solution along the further isometry $b$. By generalising this, one finds that in terms of the $\alpha=-2$  potentials T-duality acts as follows: if the  potential of one theory has no indices along $a$, then $T_a$ maps it to a  potential of the other theory with $a$ added on both the first and the second set of indices, while if the  potential of one theory has an index $a$ only along the first set of indices, it is mapped to the same potential of the other theory. It can easily be shown that these rules map the NS5-brane to all the other branes corresponding to the mixed-symmetry potentials in eq. \eqref{allDpotentials} by chains of T-dualities.

We now move to the $\alpha=-3$ branes, whose corresponding mixed-symmetry potentials are~\cite{branesandwrappingrules} 
\begin{align}
 & E_8 \  \ E_{8,2} \ \ E_{8,4} \ \ E_{9,2,1} \ \ E_{8,6} \ \ E_{9,4,1} \ \ E_{10,2,2} \ \ E_{10,4,2} \ \ E_{10,6,2} \ \ \  \ ({\rm IIB}) \nonumber \\
& E_{8,1} \ \ E_{8,3} \ \ E_{9,1,1} \ \ E_{8,5} \ \ E_{9,3,1}\  \ E_{9,5,1}\  \ E_{10,3,2} \ \ E_{10,5,2} \ \ \quad \quad \quad ({\rm IIA}) \quad , \label{allEpotentials}
\end{align}
and one can recognise in the IIB list the field $E_8$ which is the S-dual of the RR field $C_8$. We find that the T-duality rule for these potentials is the following: if the  potential of one theory has no indices along $a$, then $T_a$ maps it to a  potential of the other theory with $a$ added on three sets of indices, while if the  potential of one theory has an index $a$ only along the first set of indices, it is mapped to a potential of the other theory with  $a$ added on the first and the second set of indices. Therefore, for instance the IIB potential $E_8$ with no indices along $a$ is mapped to the IIA potential $E_{9,1,1}$ where the index $a$ is present on all three sets of indices. If instead $E_8$ has one index $a$, then it is mapped to the IIA potential $E_{8,1}$ where $a$ appears on both sets of indices. The reader can appreciate that by performing repeated T-dualities one can map the S-dual of the D7-brane (which we denote as the $7_3$-brane) to all the exotic branes corresponding to the mixed-symmetry potentials in eq. \eqref{allEpotentials}.

The rule we find actually generalises to all the other branes with more-negative $\alpha$ that occur in string theory. Given a brane with $\alpha=-n$ such that in the corresponding potential the $a$ index occurs $p$ times (in $p$ different sets of antisymmetric indices), that this is mapped by T-duality along $a$ to the brane associated to the potential in which the $a$ index occurs $n-p$ times. Schematically, we write
\begin{equation}
\alpha=-n \ : \qquad \quad \underbrace{a,a,...,a}_p \ \overset{T_a}{\longleftrightarrow} \ \underbrace{a,a, ....,a }_{n-p} \quad . \label{allbranesallalphasrule}
\end{equation}
For instance, the list of the potentials associated to the $\alpha=-4$ branes can be found in~\cite{heteroticwrappingrules}, and by looking at equations 3.6 and 3.7 and tables 9 and 10 of that paper one can show that the potentials associated to these branes are precisely related by our T-duality rules for $\alpha=-4$ in eq. \eqref{allbranesallalphasrule}. Similarly, in the same paper one can find the list of the $\alpha=-5$ potentials in  tables 11 and 12 (IIA) and 13 and 14 (IIB), and again the components of these potentials that correspond to branes are nicely related by our T-duality rules for $\alpha=-5$ in eq. \eqref{allbranesallalphasrule}. Finally, the potentials for the $\alpha=-6$ branes are given in~\cite{heteroticwrappingrules} in table 14 and equation 3.16, and the reader can check that our rule for $\alpha=-6$ in eq. \eqref{allbranesallalphasrule} works again. We have checked that the rule also works for all the other branes with more negative value of $\alpha$. In four dimensions, the lowest value of $\alpha$ is $-7$, while it is $-11$ in three dimensions~\cite{bergshoeffriccionimarrani}.

We can now move back  to the $T^6/[\mathbb{Z}_2 \times \mathbb{Z}_2]$ orientifold,  and apply the rules we have found to determine where the Chern-Simons term in eq. \eqref{SdualD7tadpole} is mapped to in the IIA theory. 
The $7_3$-branes correspond to the three components $E_{4 \, x^i y^i x^j y^j}$ (with $i \neq j$) of the 8-form potential $E_8$. 
By performing three T-dualities along the $x$ directions, this is mapped to the three components $E_{4 \, x^i y^i x^j y^j x^k, x^i x^j x^k, x^k }$  (with $i,j,k$ all different) of the potential $E_{9,3,1}$ of the IIA theory, which corresponds to a $5_3^{2,1}$-brane.\footnote{The $2,1$ denotes the fact that two isometries correspond to an index repeated twice and one isometry corresponds to an index repeated three times.}  This component has two $y$ indices, and by requiring that every component should have an even number of $y$ indices we find the additional components of $E_{9,3,1}$ that are listed in Table \ref{Ebranestable}. By mapping  these components from IIA back to  IIB, we find that in the latter theory one can include also the potentials $E_{8,4}$, $E_{9,2,1}$ and $E_{10,4,2}$, corresponding to $3^4_3$, $6^{1,1}_3$ and $5^{2,2}_3$-branes, and by including all these branes in the IIB orientifold and mapping them back to the IIA theory we arrive at a fully consistent picture in which also the allowed components of the potentials $E_{8,1}$ and $E_{10,5,2}$, corresponding to $6^1_3$ and $4^{3,2}_3$-branes, are included. The whole set of allowed potentials is summarised in Table \ref{Ebranestable}.  

\begin{table}
\begin{center}
\scalebox{1.05}{
\begin{tabular}{|c|c||c|c|}
\hline \multicolumn{2}{|c||}{IIB} & \multicolumn{2}{|c|}{IIA}\\
 \hline \rule[-1mm]{0mm}{2mm}  potential &  component & component & potential\\
 \hline \hline \rule[-2mm]{0mm}{2mm} $E_8$ & $E_{4\, x^i y^i x^j y^j}$ & $E_{4\, x^i y^i x^j y^j x^k, x^i x^j x^k, x^k }$ & $E_{9,3,1}$ \\
 \hline
  \rule[-2mm]{0mm}{2mm} $E_{8,4}$ & $E_{4\, x^i y^i x^j y^j,x^i y^i x^j y^j}$ & $E_{4\,  x^i y^i x^j y^j x^k, y^i y^j x^k, x^k }$ & $E_{9,3,1}$ \\ 
 
  \rule[-2mm]{0mm}{2mm} & $E_{4\, x^i y^i x^j y^k , x^i y^i x^j y^k } $ & $E_{4\, x^i y^i x^j y^k x^k , y^i y^k x^k  , x^k }$ & \\
 
  \cline{2-4} \rule[-2mm]{0mm}{2mm}   & $E_{4\, x^i y^i x^j x^k , x^i y^i x^j x^k }$ & $E_{4\, x^i y^i x^j x^k, y^i}$ & $E_{8,1}$ \\
  \cline{2-4} \rule[-2mm]{0mm}{2mm} & $E_{4\,  x^i y^i y^j y^k , x^i y^i y^j y^k }$ & $E_{4\, x^i y^i x^j y^j x^k y^k , y^i x^j y^j x^k y^k, x^j x^k }$ & $E_{10,5,2}$ \\
  \hline  \rule[-2mm]{0mm}{2mm} $E_{9,2,1}$ & $E_{4 \, x^i y^i x^j y^j y^k, x^i y^k, x^i}$ & $E_{4 \, y^i x^j y^j x^k y^k, x^j x^k y^k , x^k}$ & $E_{9,3,1}$ \\
 \rule[-2mm]{0mm}{2mm}  & $E_{4\,  x^i y^i x^j y^j x^k, y^i x^k , y^i}$ & $E_{4\,  x^i y^i x^j y^j x^k, x^i y^i x^j, y^i }$ &  \\

  \cline{2-4}  \rule[-2mm]{0mm}{2mm}  & $E_{4 \, x^i y^i x^j y^j x^k, x^i x^k , x^i}$ & $E_{4\, x^i x^j y^j x^k, x^j}$ & $E_{8,1}$ \\
  \cline{2-4} \rule[-2mm]{0mm}{2mm}  & $E_{4\, x^i y^i x^j y^j y^k, y^i y^k, y^i}$ & $E_{4\, x^i y^i x^j y^j x^k y^k, x^i y^i x^j x^k y^k , y^i x^k}$ & $E_{10,5,2}$ \\
  \hline   \rule[-2mm]{0mm}{2mm} $E_{10,4,2}$ & $E_{4\, x^1 y^1 x^2 y^2 x^3 y^3, x^i y^i x^j y^j, x^i y^i}$ & $E_{4\, y^i x^j y^j x^k y^k, y^i y^j x^k, y^i}$ & $E_{9,3,1}$ \\
   \rule[-2mm]{0mm}{2mm}  & $E_{4\, x^1 y^1 x^2 y^2 x^3 y^3, x^i y^j x^k y^k, x^i y^j}$ & $E_{4\, y^i x^j y^j x^k y^k, x^j y^j y^k, y^k}$ & \\
  \cline{2-4} \rule[-2mm]{0mm}{2mm}  & $E_{4\, x^1 y^1 x^2 y^2 x^3 y^3, x^i y^i x^j x^k, x^j x^k}$ & $E_{4\, x^i y^i y^j y^k , y^i}$ & $E_{8,1}$ \\

  \cline{2-4} \rule[-2mm]{0mm}{2mm}  & $E_{4\, x^1 y^1 x^2 y^2 x^3 y^3, x^i y^i y^j y^k, y^j y^k}$ & $E_{4\, x^1 y^1 x^2 y^2 x^3 y^3, y^i x^j y^j x^k y^k, y^j y^k}$ & $E_{10,5,2}$ \\
\hline
 \end{tabular}
}
\caption{\footnotesize The $\alpha=-3$ branes that can be included in order to cancel the tadpoles generated by the $P$ fluxes. In all terms, the indices $i,j,k$ are always meant to be all different.}
\label{Ebranestable}
\end{center}
\end{table}

In the next section, we will show how the inclusion of the $P$ fluxes gives rise to generalised Bianchi identities, and in particular how these give rise to specific tadpole conditions precisely for the branes that we have listed in Table \ref{Ebranestable}. This analysis will also show how various Bianchi identities that have already been considered in the literature have to be modified by the inclusion of $P$ fluxes.

\section{$P$ fluxes, Bianchi identities and tadpoles}
In this section we analyse how the $P$ fluxes modify the various Bianchi identities and tadpole conditions in our model. In particular, 
in the first subsection  we will show how the T-duality rules for the $P$ fluxes in eq. \eqref{TdualityrulesPfluxes} modify the Bianchi identities for the NS-NS fluxes, we will determine all the constraints that arise from these Bianchi identities and will comment on their solution. In the second subsection, we will then move to consider how the $P$ fluxes lead to tadpole conditions for the $\alpha=-3$ branes listed in Table \ref{Ebranestable}. 

\subsection{$P$ fluxes and NS-NS Bianchi identities}

In the absence of sources and $P$ fluxes, the NS-NS fluxes satisfy the quadratic constraints 
\begin{align}
& f_{[ab}^e H_{cd]e}=0 \nonumber \\
& Q^{ae}_{[b}H_{cd]e}+f^e_{[bc}f^a_{d]e}=0\nonumber \\
& 4 Q^{[a|e|}_{[c}f^{b]}_{d]e}+f^e_{cd}Q^{ab}_e+R^{abe}H_{cde}=0 \label{NSNSbianchinoP}\\
& R^{[ab|e|}f^{c]}_{de}+Q^{[a|e|}_d Q^{bc]}_e=0\nonumber  \\
& R^{[ab|e|}Q^{cd]}_e=0 \quad ,\nonumber  
\end{align}
which arise from their Bianchi identities~\cite{Shelton:2005cf,Aldazabal:2006up,wrase,penas,shukla1,andriot,shukla2}. By S-duality, the $Q$ flux is mapped to $P_a^{bc}$, and in \cite{Aldazabal:2006up} it was indeed shown that the second constraint in eq. \eqref{NSNSbianchinoP} is modified by the addition of the term $-P^{ae}_{[b}F_{cd]e}$, while the fourth constraint, which is the only other one that is relevant in the case of the O3 orbifold, is mapped to an equivalent quadratic constraint for $P_a^{bc}$. We now determine how the full set of quadratic constraints in eq. \eqref{NSNSbianchinoP} is modified by the inclusion of all the $P$ fluxes in the general case, and then we will analyse the particular case of the IIB/O3 and IIA/O6 models.

Our method is as follows: we first write down all the possible terms of the form $F \cdot P$ to the first equation in \eqref{NSNSbianchinoP}. These terms can only be $F_3 \wedge P_1$ and $P_1^2 \cdot F_5$ in IIB, and $P_1^1 \cdot F_4$ in IIA. Then we consider particular components of these constraints and we act on them with all possible T-dualities using eqs. \eqref{TdualityruleRRfluxes}, \eqref{TdualityruleNSfluxes} and \eqref{TdualityrulesPfluxes}. Requiring closure under T-duality fixes all the coefficients of all the possible terms of the form $F \cdot P$ that can be added to all the NS-NS quadratic constraints. Finally, we write the resulting expressions in covariant notation. The final result is that the Bianchi identities become 
\begin{align}
& 6f^{e}_{[ab}H_{cd]e}+4F_{[abc}P_{d]}+2P^{ef}_{[a}F_{bcd]ef}=0 \nonumber\\
& 3 Q^{ae}_{[b}H_{cd]e}+3f^{e}_{[bc}f^{a}_{d]e}-3P^{ae}_{[b}F_{cd]e}-P^{a,ae}F_{bcdae} +\tfrac{1}{2} P^{aefg}_{[b} F_{cd]efg}=0 \nonumber \\
& -Q^{ab}_e f^e_{cd}-4Q^{[a|e|}_{[c}f^{b]}_{d]e}-R^{abe}H_{cde} 
+2{F}_{[c}P^{ab}_{d]}+P^{a,ab}{F}_{cda}+P^{b,ab}{F}_{cdb}\nonumber \\
& \qquad \qquad +P_{[c}^{abef}{F}_{d]ef}+\tfrac{1}{2}P^{a,baef}{F}_{cdaef}-\tfrac{1}{2}P^{b,abef}{F}_{cdbef}=0 \label{NSNSBianchiIIBwithP} \\
& 3R^{[ab|e|}f^{c]}_{de}+3Q^{[ab}_eQ^{c]e}_d+P^{abce}_d {F}_e-P^{a,abce}{F}_{ade}-P^{b,abce}{F}_{bde}-P^{c,abce}{F}_{cde} \nonumber \\
& \qquad \qquad + \tfrac{1}{6} P^{a,abcefg} F_{adefg} + \tfrac{1}{6} P^{b,abcefg} F_{bdefg} + \tfrac{1}{6} P^{c,abcefg} F_{cdefg} =0 \nonumber \\
& 6R^{[ab|e|}Q^{cd]}_e+{F}_a P^{a,abcd}+{F}_b P^{b,abcd}+{F}_c P^{c,abcd}+{F}_d P^{d,abcd}+\tfrac{1}{2}P^{a,bcdaef}{F}_{aef}\nonumber \\
& \qquad \qquad +\tfrac{1}{2}P^{b,bcdaef}{F}_{bef}+\tfrac{1}{2}P^{c,bcdaef}{F}_{cef}+\tfrac{1}{2}P^{d,bcdaef}{F}_{def} =0 \nonumber 
\end{align}
in the IIB case, and 
\begin{align}
&  6f^{e}_{[ab}H_{cd]e}+4P^e_{[a}F_{bcd]e}=0 \nonumber \\
& 3Q^{ae}_{[b}H_{cd]e}+3f^{e}_{[bc}f^{a}_{d]e}-3P^{a}_{[b}F_{cd]}-P^{a,a}F_{bcda}+\tfrac{1}{2}P^{a,aef}F_{bcdaef}+\tfrac{3}{2} P^{aef}_{[b}F_{cd]ef}=0  \nonumber \\
&-Q^{ab}_e f^e_{cd}-4Q^{[a|e|}_{[c}f^{b]}_{d]e}-R^{abe}H_{cde}
+2P^{abe}_{[c}{F}_{d]e}-P^{a,abe}{F}_{cdae}-P^{b,abe}{F}_{cdbe} \nonumber \\
& \qquad \qquad  -\tfrac{1}{3} P^{abefg}_{[c} F_{d]efg} + \tfrac{1}{6} P^{a,abefg} F_{cdaefg} +\tfrac{1}{6} P^{b,abefg} F_{cdbefg} =0 \label{NSNSBianchiIIAwithP}  \\
& 3R^{[ab|e|}f^{c]}_{de}+3Q^{[ab}_eQ^{c]e}_d+ {F} P^{abc}_d-P^{a,abc}{F}_{ad}-P^{b,abc}{F}_{bd}-P^{c,abc}{F}_{cd} \nonumber \\
& \qquad \qquad -\tfrac{1}{2} P^{abcef}_d {F}_{ef}- \tfrac{1}{2} P^{a,bcaef}{F}_{daef}- \tfrac{1}{2} P^{b,bcaef}{F}_{dbef}- \tfrac{1}{2} P^{c,bcaef}{F}_{dcef}=0\nonumber \\
& 6R^{[ab|e|}Q^{cd]}_e-{F}_{de}P^{d,abcde}-{F}_{ce}P^{c,abcde}-{F}_{be}P^{b,abcde}-{F}_{ae}P^{a,abcde}=0 \nonumber 
\end{align}
in the IIA case. 

The NS-NS quadratic constraints  can be relaxed by the inclusion of sources~\cite{Villadoro:2007tb,andriot}. This obviously also applies to the constraints modified by the inclusion of $P$ fluxes in eqs. \eqref{NSNSBianchiIIBwithP} and \eqref{NSNSBianchiIIAwithP}. In particular, relaxing the first constraints (that we schematically write as $( {\rm flux} \cdot {\rm flux}  )_4 =0$) in both equations induces a charge for the NS5-brane coming from the generalised Chern-Simons term
 \begin{equation}
 \int D_6 \wedge ( {\rm flux} \cdot {\rm flux} )_4 \quad .
 \end{equation}
As we have discussed in section 3, if we now consider a particular component for $D_6$, and we perform a T-duality $T_a$ in a direction whose index $a$ is not contained in $D_6$, this is mapped to $D_{6 \, a,a}$ which is a component of the mixed-symmetry potential $D_{7,1}$. Similarly, the quadratic term in the fluxes  $({\rm flux} \cdot {\rm flux})_4$ is mapped to the component $({\rm flux} \cdot {\rm flux})^a_3$  from the second line of the  quadratic constraints in eqs. \eqref{NSNSBianchiIIBwithP} and \eqref{NSNSBianchiIIAwithP}, so that the full Chern-Simons term is mapped to
   \begin{equation}
 \int D_{6\, a,a} \wedge ( {\rm flux} \cdot {\rm flux} )^a_3 \quad . \label{CSNSNSD71}
 \end{equation}
In this expression, the $a$ index of the potential after the comma in meant to be contracted with the upstairs index of the flux term, while the other ten indices are all different. Therefore,  the three downstairs indices of the flux term are not along $a$, and in general by T-duality starting from the first Bianchi identity  one can only reach components such that the upstairs indices are all different from the downstairs ones. This means that the constraints of eqs. \eqref{NSNSBianchiIIBwithP} and \eqref{NSNSBianchiIIAwithP} (as well as the ones in eq. \eqref{NSNSbianchinoP}) are actually more than what ones gets by simply starting with the first constraints and applying T-dualities. As we will see, this point turns out to be crucial when we discuss the solutions of the quadratic constraints in the IIB and IIA orientifold models.\footnote{In our model we do not relax the NS-NS Bianchi identities because including $\alpha=-2$ branes would not be compatible with ${\cal N}=1$ supersymmetry.}

We can now study the solutions to the constraints in eqs. \eqref{NSNSBianchiIIBwithP} and \eqref{NSNSBianchiIIAwithP} for the IIB/O3 and IIA/O6 orientifolds. In the IIB/O3 case, only the second and fourth equations in \eqref{NSNSBianchiIIBwithP} are non-trivial, and can be schematically written as
\begin{equation}
(Q \cdot H_3 -P_1^2 \cdot F_3)^{a}_{bcd}=0 \label{NSNSBianchiIIBO3}
\end{equation}
and 
\begin{equation}
(Q \cdot Q-P^{1,4} \cdot F_3 )^{abc}_d=0 \quad . \label{NSNSBianchiIIBO3bis} 
\end{equation}
The relevant components of eq. \eqref{NSNSBianchiIIBO3} with $a$ different from $b,c,d$ are $(Q \cdot H_3 -P_1^2 \cdot F_3)^{x^j}_{x^i y^i y^j}$ and $(Q\cdot H_3 -P_1^2 \cdot F_3)^{y^j}_{x^i y^i x^j}$, which would induce a charge for the KK-monopoles ({\it i.e.} $5_2^1$-branes) associated to the components $D_{4\, x^k y^k x^j, x^j}$ and $D_{4\, x^k y^k y^j , y^j}$ of the mixed-symmetry potential $D_{7,1}$. By substituting the symbols given in the first columns of Tables \ref{TableRRfluxes}, \ref{TableNSfluxes} and \ref{allPfluxes} (and considering for simplicity the isotropic case), one gets the equations
\begin{align}
& \bar{a}(\bar{b}+\bar{\beta})-\bar{h}a-\bar{f}e+b\bar{h}_0+gm-q(\bar{g}+\bar{\gamma})=0 \nonumber \\
& a(b+\beta)+e(\gamma+g)+\bar{b}h_0-\bar{a}h+e_0\bar{g}+fq=0 \quad ,\label{constraintsD71braneIIB}
\end{align}
 where the notation for the isotropic fluxes is as in eq. \eqref{isotropicsuperpotential}.
Similarly, eq. \eqref{NSNSBianchiIIBO3bis}, with the index $d$ different from $a,b,c$, leads to the two components $(Q \cdot Q-P^{1,4}\cdot  F_3 )^{x^j y^j x^k}_{y^k}$ and $(Q\cdot Q-P^{1,4} \cdot F_3 )^{x^j y^j y^k}_{x^k}$, which would induce a charge for the  $5_2^3$-branes associated to the components $D_{4\, x^i y^i x^j y^j x^k, x^j y^j x^k}$ and $D_{4\, x^i y^i x^j y^j y^k, x^j y^j y^k}$ of the mixed-symmetry potential $D_{9,3}$. In the isotropic case these constraints are
\begin{align}
& -b(b+\beta)+h(\bar{b}+\bar{\beta})-f'q+e(g'+\gamma')-\bar{g}'e_0=0 \nonumber \\
& \bar{b}(\bar{b}+\bar{\beta})-\bar{h}(b+\beta)-q(\bar{\gamma}'+\bar{g}')-g'm+\bar{f}'e=0 \quad . \label{constraintsD93IIB}
\end{align}
In the IIA/O6 case, the non-trivial constraints in eq. \eqref{NSNSBianchiIIAwithP} are the second, the third and the fourth, but actually only the second and the fourth are relevant for the components such that the upstairs indices are different from the downstairs ones. The second constraint is 
\begin{equation}
 (Q\cdot H_3 + f \cdot f+P^1_1 \cdot F_2 -P^{1,1}\cdot F_4 +P_1^3 \cdot F_4+P^{1,3} \cdot F_6)^{a}_{bcd}=0 \quad ,\label{NSNSBianchiIIAO6}
 \end{equation}
 which again would induce a charge for the $5_2^1$-branes associated to the same $D_{7,1}$ components as in IIB. The constraints in this case are the second equation in \eqref{constraintsD71braneIIB} and the first in \eqref{constraintsD93IIB}. The other non-trivial constraint is 
\begin{equation}
(R \cdot  f +Q \cdot Q+ P_1^3 F_0-P^{1,3} \cdot F_2-P_1^5 \cdot F_2-P^{1,5}\cdot F_4)^{abc}_d=0 \quad , \label{NSNSBianchiIIAO6bis}
\end{equation}
which would induce a charge for the same $5_2^3$-branes as in the IIB case, leading to the first equation in  \eqref{constraintsD71braneIIB} and the second in \eqref{constraintsD93IIB}. We therefore have perfect match between the IIB and the IIA result.

The situation is different if one considers the additional non-trivial constraints that survive the orientifold projection but are not such that the upstairs and downstairs indices are all different. In the IIB case, from eq. \eqref{NSNSBianchiIIBO3} one gets the components $(Q \cdot H_3-P_1^2 \cdot F_3)^{x^j}_{x^i y^i x^j}$ and $(Q \cdot H_3-P_1^2 \cdot F_3)^{y^j}_{x^i y^i y^j}$, which in eq. \eqref{CSNSNSD71} are associated to the components $D_{4\, x^k y^k y^j, x^j}$ and $D_{4\, x^k y^k x^j , y^j}$ of the mixed-symmetry potential $D_{7,1}$, and similarly from eq. \eqref{NSNSBianchiIIBO3bis}. In the IIA case, neither eq. \eqref{NSNSBianchiIIAO6} nor eq. \eqref{NSNSBianchiIIAO6bis} lead to additional relations, while the non-trivial relations come from the  third constraint in eq. \eqref{NSNSBianchiIIAwithP}, which after the orientifold projection becomes
\begin{equation}
(-Q \cdot f-R\cdot H_3+P_1^3 \cdot F_2-P^{1,3} \cdot F_4 - P_1^5 \cdot F_4 + P^{1,5} \cdot  F_6 )^{ab}_{cd}=0 \quad . \label{NSNSBianchiIIAO6ter}
\end{equation}
What one finds is that the IIB and IIA constraints that one gets do not match, unless the additional constraints 
\begin{align} 
& q(g+\gamma)+\bar{g}e-fm+gq+e(\bar{g}+\bar{\gamma})+e_0\bar{f}=0  \label{1c}\\
& -mf'-q(g'+\gamma')+\bar{g}'e-g'q+e(\bar{g}'+\bar{\gamma}')-\bar{f}'e_0=0  \label{2cond}   \end{align} 
are satisfied.

In order to understand and solve this mismatch, we remember that the fields listed in eq. \eqref{allDpotentials}, that are 
associated to the $\alpha=-2$ branes, in the four-dimensional theory belong to representations of $SO(6,6)$. In particular, the space-filling branes correspond to a 4-form potential  $D_{4,MNPQ}$ in the ${\bf 495}$ representation~\cite{stringsolitons}. This representation not only contains the fields in  eq. \eqref{allDpotentials}, but also the potentials $D_8$, $D_{9,1}$, $D_{10}$ and $D_{10,2}$~\cite{stringsolitons}. The components of the mixed-symmetry potentials in \eqref{allDpotentials} with indices after the comma that are not parallel to any of the other indices are related by T-duality to these additional potentials.\footnote{In a group-theoretic language, all these components correspond to shorter weights of the ${\bf 495}$  representation of $SO(6,6)$ with respect to the components considered in section 3~\cite{dominantweights}.}  In particular,  the 8-form field $D_8$ is the one that together to $C_8$ and $E_8$ forms the triplet of $SL(2,\mathbb{R})$~\cite{Meessen:1998qm}. In the IIB/O3 setup, the tadpole induced by the fluxes to this potential was already considered in~\cite{Aldazabal:2006up} and arises from the Chern-Simons term
\begin{equation} 
\int D_8 \wedge (Q \cdot H_3 +P_1^2 \cdot F_3 )_2  \quad .   \label{D8tadpole} 
\end{equation} 
By imposing absence of sources for this potential, this leads to the quadratic constraint
\begin{equation} 
Q^{cd}_{[a}H_{b]cd}+P^{cd}_{[a}F_{b]cd}=0 \quad . \label{d8} 
\end{equation} 
By analysing this constraint, one finds that it implies exactly eq. \eqref{1c}.  In IIB/O3 the constraints from $D_{9,1}$ and $D_{10}$  identically vanish, while the constraint arising from the $D_{10,2}$ potential in IIB is  
\begin{equation} (P^{1,4}\cdot F_3)^{ab}=P^{a,abcd}F_{acd}+P^{b,abcd}F_{bcd}=0 \quad ,
\end{equation}
which leads exactly to the condition \eqref{2cond}. In the IIA/O6 model, the only non-trival constraint comes from $D_{9,1}$, and again it can be shown that it is perfectly compatible with all the IIB constraints.

What this analysis shows is that when the $P$ fluxes are included, one can consistently impose all the NS-NS constraints, but this also imposes for consistency that the quadratic constraints arising from the $D_8$ and $D_{10,2}$ potentials in IIB have to vanish. On the other hand, in the previous section we have shown that the $P$ fluxes also induce charges for the $\alpha=-3$ branes that can be different from zero. We now want to analyse the tadpole conditions for these branes.

\subsection{$P$ fluxes and tadpoles}
Using the T-duality rules for the $P$ fluxes and the $E$ potentials that we have found in this paper, one can determine, starting from eq. \eqref{SdualD7tadpole}, all the tadpole conditions for the $\alpha=-3$ branes listed in Table \ref{Ebranestable} in the presence of $P$ fluxes.
In the IIB/O3 theory, there are three $7_3$-branes, each orthogonal to one of the three tori $T_{(i)}^2$, corresponding to the components $E_{4\, x^jy^j x^ky^k}$ of the potential $E_8$. Denoting the number of each of these  branes as $N_{( 7_3 )_i}$, from eq. \eqref{SdualD7tadpole} one gets  \cite{Aldazabal:2006up} 
\begin{equation}
 N_{ (7_3 )_i}+\tfrac{1}{2}[-h_0\bar{f}_i+\bar{h}_0f_i+\bar{a}_jg_{ji}-a_j\bar{g}_{ji}]=0 \quad , \label{P7}
 \end{equation}
where it is understood that the index $j$ is summed.
In the IIA theory, these conditions are mapped to the conditions for the $5_3^{2,1}$-branes associated to the components $
E_{4\,x^iy^ix^jy^jx^k,x^ix^jx^k,x^k}$ of the potential $E_{9,3,1}$ (see Table \ref{Ebranestable}).  This can be shown by evaluating for these components the constraints coming from the generalised Chern-Simons term
\begin{equation}
\frac{1}{2} \int E_{ 9,abc,a} \times ( {\rm flux} \cdot {\rm flux} )^{abc,a}_1 \quad ,
\end{equation}
where the $( {\rm flux} \cdot {\rm flux} )$ term is given by 
\begin{align}
( {\rm flux} \cdot {\rm flux} )^{abc,a}_d & =-2P^{a,a[b|e|}f^{c]}_{de}+f^a_{cd}P^{c,abc}+f^a_{bd}P^{b,abc}+P^{a,a}Q^{bc}_{d}+Q^{ae}_{d}P_e^{abc}\nonumber \\ &-2P^{a[b|e}_{d}Q^{a|c]}_e+\tfrac{1}{2}P^{abcef}_{d} f^a_{ef}+P^a_{d}R^{abc}+\tfrac{1}{2}P^{a,abcef}H_{def} \quad . \label{quadratic931}
\end{align}
As we have discussed in the previous subsection, since the $5_3^{2,1}$-branes correspond to the components of the potentials such that the indices $abc$ have to be inside the first $9$ indices, this implies that for these components the index $d$ in eq. \eqref{quadratic931} differs from $a,b,c$. In particular, the components $
E_{4\,x^iy^ix^jy^jx^k,x^ix^jx^k,x^k}$  couple to the terms  $( {\rm flux} \cdot {\rm flux} )^{x^i x^j x^k, x^k}_{y^k}$, which lead precisely to the tadpole conditions equivalent to eq. \eqref{P7}.   

In section 3 we have seen that the additional $5_3^{2,1}$-branes that can be included in the IIA/O6 theory are mapped in the  IIB/O3 theory to the $3_3^4$, $6_3^{1,1}$ and $5_3^{2,2}$-branes associated to the potentials $E_{8,4}$, $E_{9,2,1}$ and $E_{10,4,2}$ respectively. 
The tadpole conditions for all these branes can be easily determined using our rules.  We write schematically the flux contributions to the tadpole conditions for all these branes as 
\begin{align}
 &P_1^2 \cdot H_3 \longleftrightarrow E_{8} \nonumber \\
 &P_1^2 \cdot  Q \longleftrightarrow E_{8,4},E_{9,2,1}\label{tadpolesforallPbranesIIBO3}\\  
 &P^{1,4} \cdot  Q \longleftrightarrow E_{10,4,2} \nonumber \quad .
\end{align}
Similarly, one can compute the tadpole conditions for the $6_3^1$ and $4_3^{3,2}$-branes in the IIA/O6 theory.  


 As an interesting application of our results, we now consider the IIB/O3 theory for the particular case in which $P^{1,4}=0$, 
and look at all the constraints related to $P_1^2 \cdot Q$ in the presence of exotic branes. From eq. \eqref{tadpolesforallPbranesIIBO3} one can see that the potential $E_{10,4,2}$ does not couple to $P_1^2$, and therefore we only have to consider, apart from $E_8$ (giving the constraint \eqref{P7}), the potentials $E_{8,4}$ and $E_{9,2,1}$. The generalised Chern-Simons term for $E_{8,4}$ is
\begin{equation}
\frac{1}{4!}\int E_{8,4} \wedge (P_1^2 \cdot Q)^4_2 \quad .\label{tadpoleE84}
\end{equation}
with $(P_1^2 \cdot Q)^{abcd}_{ef}=12 P^{[ab}_{[e}Q^{cd]}_{f]}$. We denote with $\bigcirc abcd$ the isometry directions. We find the constraints
\begin{align}
& N_{3_3^4}(\bigcirc x^jy^jx^k y^k)+\tfrac{1}{2}[g_{ii}\bar{b}_{ii}-\bar{g}_{ki}b_{ki}+\bar{f}_ih_i-\bar{g}_{ji}b_{ji}-f_i\bar{h}_i+g_{ji}\bar{b}_{ji}-\bar{g}_{ii}b_{ii}+g_{ki}\bar{b}_{ki}]=0 \nonumber \\
& N_{3_3^4}(\bigcirc y^iy^jx^ky^k)-\tfrac{1}{2}[\bar{g}_{ki}\bar{h}_j-\bar{f}_i\bar{b}_{kj}+\bar{g}_{kj}\bar{h}_i-\bar{f}_j\bar{b}_{ki}]=0 \nonumber \\
& N_{3_3^4}(\bigcirc y^ix^jx^ky^k)-\tfrac{1}{2}[-g_{ii}\bar{b}_{jj}+\bar{g}_{ji}b_{ij}+\bar{g}_{jj}b_{ii}-g_{ij}\bar{b}_{ji}]=0 \nonumber \\
& N_{3_3^4}(\bigcirc x^ix^jx^ky^k)+\tfrac{1}{2}[f_ib_{kj}-g_{ki}h_j-g_{kj}h_i+f_jb_{ki}]=0 \quad .
\end{align}
As we have already discussed in the previous subsection for the NS-NS fluxes, eq. \eqref{tadpoleE84} gives quadratic constraints also for the components that do not correspond to branes, {\it i.e.} components in which some on the downstairs indices are equal to some of the upstairs ones. These constraints are
\begin{align}
& g_{ii}\bar{b}_{ij}-\bar{g}_{ji}b_{jj}-\bar{g}_{ij}b_{ii}+g_{jj}\bar{b}_{ji}=0 \nonumber \\
& -\bar{g}_{ki}\bar{b}_{ij}+\bar{f}_ib_{jj}+\bar{g}_{ij}\bar{b}_{ki}-g_{jj}\bar{h}_i=0 \nonumber \\
& g_{ii}b_{kj}-\bar{g}_{ji}h_j-g_{kj}b_{ii}+f_j\bar{b}_{ji}=0 \nonumber \\
& -\bar{g}_{ki}b_{kj}+\bar{f}_ih_j+g_{kj}\bar{b}_{ki}-f_j\bar{h}_i=0 \label{nobraneE84} \\
& \bar{g}_{ki}\bar{b}_{jj}-\bar{f}_ib_{ij}-\bar{g}_{jj}\bar{b}_{ki}+g_{ij}\bar{h}_i=0 \nonumber \\
& -g_{ji}b_{kj}+\bar{g}_{ii}h_j+g_{kj}b_{ji}-f_j\bar{b}_{ii}=0 \nonumber \\
& g_{ji}\bar{b}_{jj}-\bar{g}_{ii}b_{ij}-\bar{g}_{jj}b_{ji}+g_{ij}\bar{b}_{ii}=0 \quad .\nonumber
\end{align}

For the $E_{9,2,1}$ potential, the generalised Chern-Simons term has the form
\begin{equation}
\frac{1}{2}\int E_{9,2,1} \wedge (P_1^2 \cdot Q)^{2,1}_1 \quad ,
\end{equation}
with $(P_1^2\cdot Q)^{ab,c}_d=  \frac{1}{3} (-P^{be}_d Q^{ac}_e-2P^{ce}_dQ^{ab}_e-P^{ae}_dQ^{cb}_e+Q^{ae}_d P^{cb}_e+Q^{be}_dP^{ac}_e+2Q^{ce}_dP^{ab}_e)$. 
The exotic branes are the $6_3^{1,1}$-branes, and denoting with $//abc$ the internal directions wrapped by the branes, and with $\bigcirc d ,\bigcirc e$ the isometries corresponding to the index $d$ repeated twice and the index $e$ repeated three times, the constraints are
\begin{align}
& N_{6_3^{1,1}}(//x^jy^jy^i, \bigcirc x^k ,\bigcirc x^i)-\tfrac{1}{2}[f_kb_{jj}-g_{jk}h_j-h_kg_{jj}+b_{jk}f_j]=0\nonumber \\
& N_{6_3^{1,1}}(// x^ix^jy^j,\bigcirc x^k, \bigcirc y^i)-\tfrac{1}{2}[g_{ik}\bar{b}_{kj}-\bar{g}_{kk}b_{ij}-b_{ik}\bar{g}_{kj}+\bar{b}_{kk}g_{ij}]=0 \nonumber \\
& N_{6_3^{1,1}}(//y^ix^jy^j, \bigcirc y^k, \bigcirc x^i)+\tfrac{1}{2}[-g_{kk}\bar{b}_{ij}+\bar{g}_{ik}b_{kj}+b_{kk}\bar{g}_{ij}-\bar{b}_{ik}g_{kj}]=0 \nonumber \\
& N_{6_3^{1,1}}(// x^ix^jy^j,\bigcirc y^k, \bigcirc y^i)+\tfrac{1}{2}[-\bar{g}_{jk}\bar{h}_j+\bar{f}_k\bar{b}_{jj}+\bar{b}_{jk}\bar{f}_j-\bar{h}_k\bar{g}_{jj}]=0 \quad .
\end{align}
As in the previous case, we must also consider the quadratic constraints that do not correspond to branes. These are
\begin{align}
& g_{kk}\bar{b}_{kj}-\bar{g}_{ik}b_{ij}-\bar{g}_{jk}b_{jj}+\bar{f}_kh_j-b_{kk}\bar{g}_{kj}+\bar{b}_{ik}g_{ij}+\bar{b}_{jk}g_{jj}-\bar{h}_kf_j=0\nonumber \\
& -f_k\bar{h}_j+g_{jk}\bar{b}_{jj}+g_{ik}\bar{b}_{ij}-\bar{g}_{kk}b_{kj}+h_k\bar{f}_j-b_{jk}\bar{g}_{jj}-b_{ik}\bar{g}_{ij}+\bar{b}_{kk}g_{kj}=0\nonumber 
\\
& -g_{kk}\bar{h}_j+\bar{g}_{ik}\bar{b}_{jj}+\bar{g}_{jk}\bar{b}_{ij}-\bar{f}_kb_{kj}+b_{kk}\bar{f}_j-\bar{b}_{ik}\bar{g}_{jj}-\bar{b}_{jk}\bar{g}_{ij}+\bar{h}_kg_{kj}=0\nonumber \\
& f_k\bar{b}_{kj}-g_{jk}b_{ij}-g_{ik}b_{jj}+\bar{g}_{kk}h_j-h_k\bar{g}_{kj}+b_{jk}g_{ij}+b_{ik}g_{jj}-\bar{b}_{kk}f_j=0\nonumber \\
& -2\bar{g}_{jk}b_{jj}+2\bar{f}_kh_j-g_{kk}\bar{b}_{kj}+\bar{g}_{ik}b_{ij}+b_{kk}\bar{g}_{kj}-\bar{b}_{ik}g_{ij}+2\bar{b}_{jk}g_{jj}-2\bar{h}_kf_j=0\nonumber \\
& 2g_{ik}\bar{b}_{ij}-2\bar{g}_{kk}b_{kj}+f_k\bar{h}_j-g_{jk}\bar{b}_{jj}-h_k\bar{f}_j+b_{jk}\bar{g}_{jj}-2b_{ik}\bar{g}_{ij}+2\bar{b}_{kk}g_{kj}=0\nonumber \\
& 2\bar{g}_{jk}\bar{b}_{ij}-2\bar{f}_kb_{kj}+g_{kk}\bar{h}_j-\bar{g}_{ik}\bar{b}_{jj}-b_{kk}\bar{f}_j+\bar{b}_{ik}\bar{g}_{jj}-2\bar{b}_{jk}\bar{g}_{ij}+2\bar{h}_kg_{kj}=0\nonumber \\
& -2g_{ik}b_{jj}+2\bar{g}_{kk}h_j-f_k\bar{b}_{kj}+g_{jk}b_{ij}+h_k\bar{g}_{kj}-b_{jk}g_{ij}+2b_{ik}g_{jj}-2\bar{b}_{kk}f_j=0\nonumber \\
& g_{kk}b_{jj}-\bar{g}_{ik}h_j-b_{kk}g_{jj}+\bar{b}_{ik}f_j=0\nonumber\\
& \bar{g}_{jk}\bar{b}_{kj}-\bar{f}_kb_{ij}-\bar{b}_{jk}\bar{g}_{kj}+\bar{h}_kg_{ij}=0\nonumber \\
& -f_k\bar{b}_{ij}+g_{jk}b_{kj}+h_k\bar{g}_{ij}-b_{jk}g_{kj}=0\nonumber \\
& -g_{ij}\bar{h}_j+\bar{g}_{kk}\bar{b}_{jj}+b_{ik}\bar{f}_j-\bar{b}_{kk}\bar{g}_{jj}=0 \quad .
\end{align}

Exactly as we have discussed in the previous subsection, what we have determined is not yet the full set of constraints. Indeed, the four-dimensional $\alpha=-3$ space-filling branes correspond to the 4-form potential $E_{4, MN \dot{\alpha}}$ belonging to the `tensor-spinor' ${\overline{\bf 1728}}$ representation of $SO(6,6)$. Together with the potentials in eq. \eqref{allEpotentials} that are associated to the branes, there are additional potentials that must be included in order to generate the whole four-dimensional representation. Focusing on the IIB/O3 model,
it turns out that in order to get all the possible constraints for $P_1^2 \cdot Q$ one has to introduce also the  fields $E_{9,3}$ and $E_{10,2}$. These potentials  occur in the IIB decomposition of $E_{11}$ and correspond to roots with  zero and negative squared length respectively  \cite{Kleinschmidt:2003mf}.  In \cite{axel} it was shown that only the $E_{11}$ roots with positive squared length are associated to branes.

The constraint corresponding to $E_{9,3}$ is 
\begin{equation}
Q^{[ab}_e P^{c]e}_d+ P^{[ab}_e Q^{c]e}_d=0 \quad , 
\end{equation}
and  has already been proposed in \cite{Aldazabal:2006up}.  
 In components one gets
\begin{align}
& -b_{jj}\bar{g}_{jk}+\bar{b}_{kj}g_{kk}+h_j\bar{f}_k-b_{ij}\bar{g}_{ik}-g_{jj}\bar{b}_{jk}+f_j\bar{h}_k+\bar{g}_{kj}b_{kk}-g_{ij}\bar{b}_{ik}=0 \nonumber \\
& \bar{b}_{ij}g_{ik}-\bar{h}_jf_k-b_{kj}\bar{g}_{kk}+\bar{b}_{jj}g_{jk}+\bar{g}_{ij}b_{ik}-g_{kj}\bar{b}_{kk}-\bar{f}_jh_k+\bar{g}_{jj}b_{jk}=0\nonumber \\
& -b_{jj}g_{ik}+\bar{b}_{kj}f_k+h_j\bar{g}_{kk}-b_{ij}g_{jk}-g_{jj}b_{ik}+f_j\bar{b}_{kk}+\bar{g}_{kj}h_k-g_{ij}b_{jk}=0\nonumber \\
& \bar{b}_{ij}\bar{g}_{jk}-\bar{h}_jg_{kk}-b_{kj}\bar{f}_k+\bar{b}_{jj}\bar{g}_{ik}+\bar{g}_{ij}\bar{b}_{jk}-g_{kj}\bar{h}_k-\bar{f}_jb_{kk}+\bar{g}_{jj}\bar{b}_{ik}=0 \quad .
\end{align}
From the multiplicity analysis of $E_{11}$ one can show that there are actually three independent 
$E_{10,2}$ potentials \cite{Kleinschmidt:2003mf}: with respect to the $SL(2,\mathbb{R})$ symmetry of the IIB theory, one belongs to the  triplet that also contains $D_{10,2}$, while the other two are singlets. If only $P_1^2$ and $Q$ fluxes are turned on, the constraint arising from the triplet vanishes, while from the two singlets one gets
\begin{equation}
Q^{e[a}_f P^{b]f}_e =0 \quad ,
\end{equation}
which in components gives
\begin{align}
& -b_{kk}\bar{g}_{kj}+h_k\bar{f}_j+\bar{b}_{ik}g_{ij}-b_{jk}\bar{g}_{jj}-b_{jj}\bar{g}_{jk}+h_j\bar{f}_k+\bar{b}_{ij}g_{ik}-b_{kj}\bar{g}_{kk} \nonumber \\
& + \bar{b}_{jk}g_{jj}-b_{ik}\bar{g}_{ij}-\bar{h}_kf_j+\bar{b}_{kk}g_{kj}+\bar{b}_{kj}g_{kk}-b_{ij}\bar{g}_{ik}-\bar{h}_jf_k+\bar{b}_{jj}g_{jk}=0 \quad .
\end{align}

One can  solve the whole set of equations that we have determined. In particular, in the 
isotropic case and without localised sources, a simple solution is 
\begin{equation}
 g=\bar{g}=\gamma=\bar{\gamma}=f=\bar{f} \quad  \text{and} \quad  b=\bar{b}=\beta=\bar{\beta}=h=\bar{h} \quad .
 \end{equation} 
A further investigation of the solutions that one can find, as well as a more general analysis of the constraints when all the allowed fluxes and branes are turned on, is beyond the scope of this paper.

\section{Conclusions}

In this paper we have derived the T-duality transformation rules for the $P$ fluxes, and this allowed us to write their contribution  to the superpotential for the $T^6/[\mathbb{Z}_2 \times \mathbb{Z}_2 ]$ IIB/O3 and IIA/O6 orientifold models. The IIB/O3 orientifold result reproduces the general expression found in \cite{Aldazabal:2010ef} as far as these fluxes are concerned. The $P$ flux contribution to the superpotential in the IIB/O3 case amounts to a term proportional to $ST$ and a term proportional to $T^2$, both multiplying cubic polynomial in $U$. The IIA/O6 superpotential has the same expression with $U$ and $T$ interchanged.  The $P$ fluxes also modify the Bianchi identities for the NS-NS fluxes, and generate tadpoles for the $\alpha=-3$ branes. To compute the tadpole conditions, one has to determine how these branes transform under T-duality, and we have achieved this in this paper by determining a 
universal T-duality rule for all the branes in string theory. 

The expressions for the NS-NS Bianchi identities with $P$ fluxes included are given in eqs. \eqref{NSNSBianchiIIBwithP} and \eqref{NSNSBianchiIIAwithP}. These conditions are only compatible with the duality between the IIB/O3 and the IIA/O6 theory if additional constraints are imposed, which  in particular would  imply that  a  source for the $D_8$ potential that is dual to the dilaton can not be included.\footnote{This source is commonly known as the $I7$-brane in the literature \cite{Aldazabal:2006up}.} As far as the Bianchi identities for the $P$ fluxes are concerned, we claim that they can be consistently relaxed for all the components that generate charges for the exotic branes listed in Table \ref{Ebranestable}, while they are still satisfied for the other components. We have used the brane classification carried out in \cite{Bergshoeff:2010xc,stringsolitons,bergshoeffriccionimarrani,branesandwrappingrules}, and in particular we associate specific components of  mixed-symmetry potentials to exotic branes. 

This analysis can be extended in different directions. First of all, one can study in more detail the solutions of the tadpole conditions that we find, and plug them in the superpotential in order to minimise the scalar potential. Moreover,  one can complete the analysis by including all the possible fluxes and branes in the model, both in the IIB and the IIA setup. This would generally lead to a superpotential  which is cubic in both the $U$ and $T$ moduli~\cite{Aldazabal:2006up}, with more general tadpole conditions than what we find in this paper. This would be of interest in the context of moduli stabilisation and more generally for phenomenological applications.

The superpotential of the IIA/O6 theory given  in eq. \eqref{Waallpfluxes} was obtained in this paper by simply requiring the matching with IIB using the mapping dictated by the T-duality transformation rules that we have found. It would be interesting to understand whether this expression has a validity for  generic IIA/O6 Calabi-Yau compactifications. This would give the 
 equivalent to the analysis carried out in  \cite{Aldazabal:2010ef} for IIA as far as $P$ fluxes are concerned. One could then try to extend this analysis for all the fluxes of the IIA theory.

Obviously, it would be of extreme interest to get any understanding of the dynamics of the exotic branes that according to our analysis can be consistently introduced to cancel the charges induced by the $P$ fluxes. We hope that this work could in principle shed some light on this crucial issue.

\vskip 1cm

\section*{Acknowledgements}
FR would like to acknowledge hospitality of the Galileo Galilei Institute for Theoretical Physics and thank the organisers of  the  `Supergravity: what next?' workshop, where part of this work has been carried out, for creating a stimulating atmosphere. We would like to thank  G. Pradisi for discussions at various stages of this project. FR would like to thank  G. Dibitetto for discussions on tadpole conditions.

\end{document}